\documentclass{sig-alternate}
\pdfoutput=1
\newif\ifwww
\ifwww
\permission{Copyright is held by the International World Wide Web Conference Committee (IW3C2). IW3C2 reserves the right to provide a hyperlink to the author's site if the Material is used in electronic media.}
\conferenceinfo{WWW'14,}{April 7--11, 2014, Seoul, Korea.}
\copyrightetc{ACM \the\acmcopyr}
\crdata{978-1-4503-2744-2/14/04. \\ http://dx.doi.org/10.1145/2566486.2568046}
\else
\makeatletter
\def\@copyrightspace{\relax}
\makeatother
\fi

\usepackage{color}
\PassOptionsToPackage{hyphens}{url}

\usepackage[pdftex]{hyperref}
\usepackage{hyperref} 
\def\authortext{H. T. T. Truong, E. Lagerspetz,  P. Nurmi, A. J. Oliner, S. Tarkoma, N. Asokan, S. Bhattacharya}

\def\keywordstext{mobile malware; infection rate; Android; malware detection}

\def\titletext{The Company You Keep: Mobile Malware Infection Rates and Inexpensive Risk Indicators}

\hypersetup{
pdftitle={\titletext},
pdfauthor={\authortext},
pdfkeywords={\keywordstext},
bookmarksnumbered,
colorlinks,
citecolor=black,
filecolor=black,
linkcolor=black,
urlcolor=black,
breaklinks=true
}

\usepackage[plain]{algorithm}
\usepackage{breakurl}
\usepackage{graphicx}
\usepackage{fancyhdr}
\usepackage{xspace}
\usepackage{pseudocode}
\usepackage{tabularx}
\usepackage[figuresright]{rotating}
\usepackage{changepage}
\usepackage[table]{xcolor}
\usepackage{hhline,balance}
\usepackage{datetime}
\usepackage{caption}
\usepackage{subcaption}
\usepackage{stfloats}
\usepackage{multirow}
\usepackage{booktabs}
\definecolor{mygray}{rgb}{0.9,0.9,0.9}

\newcommand{\squishlist}{
 \begin{list}{$\bullet$}
  { \setlength{\itemsep}{0pt}
     \setlength{\parsep}{3pt}
     \setlength{\topsep}{3pt}
     \setlength{\partopsep}{0pt}
     \setlength{\leftmargin}{1.5em}
     \setlength{\labelwidth}{1em}
     \setlength{\labelsep}{0.5em} } }

\newcommand{\squishlisttwo}{
 \begin{list}{$\bullet$}
  { \setlength{\itemsep}{0pt}
     \setlength{\parsep}{0pt}
    \setlength{\topsep}{0pt}
    \setlength{\partopsep}{0pt}
    \setlength{\leftmargin}{2em}
    \setlength{\labelwidth}{1.5em}
    \setlength{\labelsep}{0.5em} } }

\newcommand{\squishend}{
  \end{list}  }

\newcommand{\caratapp}{Carat\xspace}

\newcommand{\caratrefsonly}{\cite{caratSensys2013}}
\newcommand{\caraturl}{http://carat.cs.berkeley.edu/}
\newcommand{\predictor}{indicator}

\newcommand{\McAfeeinfectionrate}{0.28\%\xspace}
\newcommand{\MobSandinfectionrate}{0.26\%\xspace}
\newcommand{\infectionrate}{0.51\%\xspace}

\newcommand{\dcpv}{$<$dc,p,v$>$\xspace}
\newcommand{\dcp}{$<$dc,p$>$\xspace}
\newcommand{\dconly}{$<$dc$>$\xspace}
\newcommand{\dcvalue}{\textbf{dc}\xspace}
\newcommand{\pvalue}{\textbf{p}\xspace}
\newcommand{\vvalue}{\textbf{v}\xspace}

\DeclareCaptionType{copyrightbox}

\title{\titletext}

\numberofauthors{1}
\author{
\alignauthor Hien Thi Thu Truong \textsuperscript{\o}, Eemil Lagerspetz \textsuperscript{\o$\S$}, Petteri Nurmi \textsuperscript{\o$\S$}, Adam J.\ Oliner \textsuperscript{$\dag$}, \\Sasu Tarkoma \textsuperscript{\o$\S$}, N. Asokan \textsuperscript{$\ddag$\o}, Sourav Bhattacharya \textsuperscript{\o$\S$} \\\vspace{1em}
\affaddr{\textsuperscript{\o} University of Helsinki, Gustaf H\"{a}llstr\"{o}min katu 2b, 00560 Helsinki, Finland \\
\textsuperscript{$\S$} Helsinki Institute for Information Technology HIIT, Gustaf H\"{a}llstr\"{o}min katu 2b, 00560 Helsinki, Finland \\
\textsuperscript{$\ddag$} Aalto University, Otakaari 4, 02150 Espoo, Finland \\
\textsuperscript{$\dag$} University of California at Berkeley, CA 94720, USA}\\\vspace{1em}
\email{\o\ first.last@cs.helsinki.fi, $\ddag$ asokan@acm.org, $\dag$ oliner@eecs.berkeley.edu}
}

\begin{document}

\maketitle

\begin{abstract}
  There is little information from independent sources in the public
  domain about mobile malware infection rates. The only previous
  independent estimate (0.0009\%)~\cite{charles_lever_core_2013}, was
  based on indirect measurements obtained from domain-name resolution
  traces.  In this paper, we present the first independent study of
  malware infection rates and associated risk factors using data
  collected directly from over 55,000 Android devices. We find that
  the malware infection rates in Android devices estimated using two
  malware datasets (\McAfeeinfectionrate and \MobSandinfectionrate),
  though small, are significantly higher than the previous independent
  estimate.  \ifwww
%% this sentence is somewhat redundant now given the next paragraph.
%% So take it out of the WWW version.
\else
Using
  our datasets, we investigate how {\predictor}s extracted
  inexpensively from the devices correlate with malware infection. 
\fi

% As
%   part of the analysis, we examine whether it is possible to cheaply
%   predict likely malware infection on a mobile device by examining
%   \emph{only} a set of reliable identifiers for the applications that
%   the user has run,

  Based on the hypothesis that some application stores have a greater
  density of malicious applications and that advertising within
  applications and cross-promotional deals may act as infection
  vectors, we investigate whether the set of applications used on a
  device can serve as an {\predictor} for infection of that device.
  Our analysis indicates that, while not an accurate indicator of
  infection by itself, the application set does serve as an
  inexpensive method for identifying the pool of devices on which more
  expensive monitoring and analysis mechanisms should be deployed.
  Using our two malware datasets we show that this {\predictor}
  performs up to about five times better at identifying infected devices than
  the baseline of random checks.  Such {\predictor}s can be used, for
  example, in the search for new or previously undetected malware. It
  is therefore a technique that can complement standard malware
  scanning.  Our analysis also demonstrates a marginally significant
  difference in battery use between infected and clean devices.

\end{abstract}

\ifwww
% ACM 1998 classification
\else
% new classification
\fi

%% put a watermark on the full version
%\ifsubmission
%\else
% \SetWatermarkText{------------------------------------------    Draft:
%   please do not redistribute}
% \SetWatermarkLightness{ 0.95 }
% \SetWatermarkScale{1}
% \SetWatermarkAngle{ -45 }
% \SetWatermarkFontSize{2cm }
%\begin{center}
%\textbf{Version of November 14, 2013}
%\textbf{Version of \today}
%\end{center}
%\vspace{3cm}
%\fi
%\vspace{2cm}
\section{Introduction}
\label{sec:intro}
How prevalent is mobile malware? There has been a steady stream of
popular news articles and commercial press releases asserting that the
mobile malware problem, especially on the Android platform, is
dire~\cite{brian_krebs_mobile_2013_wurl,damballa_labs_damballa_2011_wurl,nqmobile_mobile_2013_wurl}.
For example, a recent press release~\cite{nqmobile_mobile_2013_wurl}
claims that 32.8 million Android devices were infected in 2012, which,
given the estimated 750 million Android
devices~\cite{larry_page_update_2013_wurl}, works out to an infection
rate of 4.3\%.  Lookout mobile reported that the global malware
infection rate (which they call ``likelihood of encountering
application-based mobile threats'') for Lookout users is 2.61\%,
consisting of various types of threats ranging from adware (most
prevalent at 1.6\%) to spyware (least prevalent at 0.1\%) in their
latest published estimate (June 2013)~\cite{Lookout13}.

Some reports, however, speculate the opposite---that actual infections
in the wild are rare, especially in the
West~\cite{robert_mcgarvey_threat_2013}.  There are other indications
that the malware problem in mobile devices may indeed not be severe.
Google Play, and other legitimate application markets, actively use
malware scanners to detect and remove malware.  Kostiainen \textit{et
al.}~\cite{DBLP:conf/codaspy/KostiainenREA11} describe the widespread
availability of hardware and software platform security mechanisms on
mobile devices, which could make them more robust against malware
compared to traditional personal computers.
\ifwww
\else
Husted \textit{et
al.}~\cite{husted_nathaniel_why_2011} use analytical modeling to
conclude that mobile-to-mobile infection of malware is not a
significant threat.
\fi

The research community has focused primarily on analyzing malware to
detect if a given software package is malware or not and studying how
malware may spread. The only independent research so far to address
the question of malware infection rate was by Lever~\textit{et
al.}~\cite{charles_lever_core_2013} which used an indirect method of
inferring infection by analyzing domain name resolution queries. They
concluded that the infection rate in the United States is less than
0.0009\% (c.f.\ 2.61\% and 4.3\% above). What accounts for this
disparity?

There has been little direct measurement of malware infection rates in
the wild by \textit{independent sources}. In order to address this
dearth, we carry out a large-scale study by gathering data from tens
of thousands of Android devices in the wild.  We instrumented
\caratapp~\caratrefsonly{}, a popular Android application, to collect information about the identities of applications run on devices.  We used
the resulting dataset to look for the presence of malware based on
three different Android malware datasets.

Furthermore, we used our dataset to study the likely influence of
several potential risk factors. First, hypothesizing that malware may
be a battery hog, we compared the distribution of battery lifetime
estimates between the sets of infected and clean devices.  Second,
following the adage ``you are the company you keep,'' we hypothesized
that the set of applications used on a device may predict the
likelihood of the device being classified as infected in the future.
Our intuition behind this hypothesis is the possibility that the set
of applications used on a device is indicative of user behavior, such
as which application stores they use.

Our intent is to identify a small set of ``vulnerable'' devices (from
among a large population) based on {\predictor}s that can be measured
inexpensively on the device. More extensive monitoring and analysis
techniques can then be deployed on that subset of devices.  Such an
approach can complement and support anti-malware tools in several
ways.  First, since our technique focuses on estimating the infection
susceptibility of a device (rather than on classifying whether a given
package is malware), it may help in the search for \textit{previously
undetected malware} by allowing anti-malware vendors to focus on the
applications present on the small set of vulnerable devices.  Second,
it can provide an early warning to enterprise administrators,
especially in those enterprises with a Bring Your Own Device policy to
identify vulnerable users for whom they can provide additional help or
training.

The contributions of this paper are:
\begin{itemize}
\item \textbf{The first independent, directly measured
    estimate of mobile malware infection rate}.
% We believe this is the first \emph{independent}
%   large-scale
%   study of mobile malware infection rates using data gathered \emph{directly}
%   from the devices. 
  Taking a conservative approach to identifying malware, we used two
  malware datasets to find infection rates of \McAfeeinfectionrate
  and \MobSandinfectionrate, which is significantly more than the
  previous independent estimate~\cite{charles_lever_core_2013}
  (Section~\ref{sec:infectionrates}).

  % A closer inspection reveals that some of the most
  % frequently detected ``malware'' are rootkits and ad libraries that
  % may have been installed intentionally by the users.  If we discount
  % such packages from the malware dataset, the infection rate is
  % TODO\%, which is (TODO: closer to or still bigger than?) the
  % estimate in~\cite{charles_lever_core_2013}.

  % TODO: check what we can say about the estimate of
  % ~\cite{nqmobile_mobile_2013_wurl}.  Does the dc-only match, after
  % pruning obvious leaked keys, come closer?

%% \item \textbf{Quantification of the effect of infection on battery
%% use}. We show that infected devices (labeled according to the two
%% malware datasets) have a lower expected battery life
%%   (by 1.3 and 0.4 hours respectively) than clean devices and the difference is
%%   marginally significant
%%   (Section~\ref{subsec:predictions:energy}).
%% % ADAM these contributions should all have forward pointers -- DONE

\item \textbf{A lightweight technique to detect susceptibility of a
    device to infection}. We propose a new approach to quantify
  susceptibility of a device to malware infection based only on a set
  of identifiers representing the applications used on a
  device. Compared with random selection, our method---which requires
  only lightweight instrumentation---shows a five-fold improvement
  (Section~\ref{subsec:predictions:applications}).

\end{itemize}

We begin by discussing the background of our data collection methodology
(Section~\ref{sec:background}), and the datasets we use
(Section~\ref{sec:datasets}).  We then present infection rates
inferred from the datasets (Section~\ref{sec:infectionrates}) before
analyzing potential infection indicators and explaining the method for
building detection models (Section~\ref{sec:predictions}).  We then
discuss the big picture (Section~\ref{sec:discussion}) and provide an
account of related work (Section~\ref{sec:relatedwork}), before
concluding (Section~\ref{sec:conclusion}).

%
%\newpage
\section{Background}
\label{sec:background}
\ifwww
\else
In this section, we describe the background: our approach for unique
identification of Android packages and a summary of \caratapp{} which
we use as the vehicle for our data collection.
\fi

%\newpage
\subsection{Identifying Android Packages}
\label{subsec:background:android-packageid}

An Android package is identified by a Java language-style package name
that is intended to be globally unique. To ensure uniqueness, the
recommended naming scheme is for developers to base the package name
on an Internet domain name that they own. An example of such a package
name is \texttt{com.facebook.katana}, which is the official Facebook
application. An Android device enforces uniqueness of package names
within it. Google Play, the official Android application market, also
enforces unique package names across the application market.  However,
using unique package names is mere convention, and nothing prevents
any developer from generating a package with any name. For example,
the name
\texttt{com.facebook.katana} is used in many malware packages.
Zhou~et al.~\cite{zhou_dissecting_2012} demonstrated that
repackaging popular Android applications with a malicious payload is a
favorite technique of Android malware authors. Therefore the package
name alone is not a reliable identifier for uniquely identifying an
Android package.

All Android packages must be signed. Google recommends that each
developer have a long-term key pair for signing all their
applications\footnote{\url{http://developer.android.com/tools/publishing/app-signing.html}}. Developer
signing keys are often used with self-signed certificates. The
certificate(s) (\emph{devcert}) can be extracted from the certificate
files in the META-INF directory within an Android package or by an
application from within an Android device from the (misleadingly
named) \texttt{Signature} field of the
\texttt{PackageInfo} object via the \texttt{PackageManager} class. A
package may be signed with multiple keys. We use the
terms \emph{package} and \emph{application} interchangeably.

Each Android package also has version information in the form of a
numeric ``versionCode'' (integer) and ``versionName'' (string,
intended to be displayed to users).  A developer is required to update
the versionCode whenever they release a new version of a package
because versionCode is used to decide if a package on a device needs
to be updated.  Therefore instead of just the package name we can use
the combination of the package name, devcert, and versionCode as a
reliable unique identifier for Android packages. We use the
tuple \dcpv to identify a package, where:

\squishlist
\begin{itemize}

\item \dcvalue: a statistically unique ID for the developer
  (devcert), generated as a SHA1 hash of devcert;

\item \pvalue: the Android package name, extracted from \texttt{AndroidManifest.xml}, or from the system process list on the device;

\item\vvalue: the versionCode the package, also obtained from 
  \texttt{AndroidManifest.xml}.
\end{itemize}
\squishend

% Carat background

\subsection{\caratapp Application}
\label{subsec:background:carat}

We collected data using a modified version of \caratapp~\caratrefsonly, a mobile application that runs on stock Android and iOS devices. \caratapp{} uses energy-efficient, non-invasive instrumentation to intermittently record the state of the device, including the battery level and process list, and uses this information to recommend actions that help users improve the device's battery life.
The application is deployed to over 650,000 devices (41\% Android) and is publicly available from Google's Play Store and Apple's App Store. In this paper, we discuss data from the Android version.

\caratapp{} records several pieces of information when the device
battery level changes (sampling is triggered by the \texttt{BATTERY\_CHANGED}
Intent), including:
\squishlist
%\item CPU usage from two point measurements of /proc/cpuinfo;
\item Process list with package names, human-readable names, application type
  (system application or not), its priority and process id;
%\item Screen brightness (0-255) and whether brightness is set to automatic or manual
\item Network type (Wi-Fi, cellular, or disconnected)
%\item Mobile network type (1xRTT, CDMA, EDGE, EVDO\_0/\_A, GPRS, HSDPA, HSPA, ...)
%\item Roaming on/off
%\item Mobile data on/off
%\item Mobile data activity in/out/inout/idle
%\item Wifi enabled/disabled
%\item Wifi signal strength
%\item Wifi link speed
\item Battery information (level, health, voltage, temperature and status)
% \item Battery level (between 0 and 1);
% \item Battery health (good, dead, cold, or hot);
% \item Battery voltage in Volts;
% \item Battery temperature in Celsius;
% \item Battery status (charging or discharging); and
%\item Charger type (USB or AC).
%\item Memory used/free/active/inactive
%\item distance travelled since last sample (calculated from coarse location)
\squishend

\caratapp{} sends the collected data to a collaborative analysis backend
running on a cloud-based cluster of Amazon EC2 instances. The analysis identifies applications that use anomalously large amounts of energy, either relative to other applications or to other instances of the same application on other devices, and reports any discovered anomalies back to the affected devices.

We chose \caratapp{} because of its high visibility,
large user base, and the willingness of the \caratapp{} development team
to let us instrument \caratapp{} as
needed. Carat has an international user base,
representative of current mobile device users.
Roughly 36\% of Carat users are based in North America,
24\% in Asia, 23\% in Europe, and 10\% in the Middle East.
Section~\ref{subsec:datasets:carat} explains how we
changed the application to help identify potential infection. 
%% Comment this out; This belongs to the discussion section.  Also
%% we want to emphasize the ability to pinpoint device infection,
%% rather than specific malicious applications.
% The results of
% this paper can further improve \caratapp{} by distinguishing malicious
% applications from legitimate ones, and warning users about malicious applications.

%
\section{Datasets}
\label{sec:datasets}
\ifwww
\else
In this section we describe in detail the different datasets used
throughout the paper: the Carat dataset and the three
malware datasets used in our experiments. We also discuss ethical
considerations of collecting data from real users.
\fi
\subsection{\caratapp Dataset}
\label{subsec:datasets:carat}
We modified the open-source\footnote{\url{\caraturl}} \caratapp
application to record the unique developer IDs \dcvalue, as described
in Section~\ref{subsec:background:android-packageid}.
\ifwww
\else
(SHA-1 hash of the developer certificates of
each package from the \texttt{PackageInfo} object.)
\fi
The \caratapp development team published our modified version
of \caratapp{} on March 11, 2013 and provided us with data collected
from that date until October 15, 2013. There are $55,278$ Android
devices that were updated to the new version during the data
collection period and reported package identification information.

Each device running Carat is identified by a Carat ID which is a
statistically unique, \caratapp{}-specific device identifier, computed
by applying SHA-1 to a concatenation of available device identifiers
(e.g., IMEI and WiFi MAC address) and the \caratapp{} installation
time. When \caratapp{} takes a sample, it walks the process list and
generates a record for each running application. Package information
is extracted on-device directly from \texttt{PackageInfo}. In addition
to Carat ID and package identifiers, \caratapp{} also records the
\textit{translated name} of the package (the human-readable string
identifying the application to users), the permissions of the package,
and the \textit{timestamp} when the package was recorded by
\caratapp{}. The additional information is used for a different
analysis, described in Section~\ref{sec:predictions}.  An entry in the \caratapp{} dataset,
summarized in Table~\ref{tab:caratdataset}, is of the form
\textbf{$<$ID, dc, p, v$>$}.
%%%%
\begin{table}[!htb]
  \centering
  \begin{tabular}{|l|r|}
      \toprule
      Type &                                        Count   \\
      \midrule
      distinct devices                       & 55,278  \\
      unique package names                   & 64,916 \\
      unique devcerts (dc)                   & 41,809 \\ %21,486  \\
      unique \dcp tuples                     & 83,226 \\ %48,980  \\
      unique \dcpv tuples                    & 192,080 \\
      total unique records                   & 5,358,819 \\
      \bottomrule
    \end{tabular}
  
  \caption{Summary of the \caratapp{} dataset.\label{tab:caratdataset}}
\end{table}
%%%%
%% Each \caratapp record also has additional information that we do not
%% fully use in this paper: \squishlist
%% \item \textbf{time}, the UNIX-style timestamp when the sample was
%%   generated;
%% \item \textbf {transname}, the human-readable string identifying the
%%   application to users, from \texttt{PackageInfo}.
%% \item \textbf {permission}, the vector of permission of a package.
%%   \squishend We explain how these can be used in
%%   Section~\ref{sec:predictions}.  For privacy reasons, \caratapp{}
%%   does not collect any personally identifying information about the
%%   user.
%%%%%%
The authors of \caratapp provide more details about the privacy
protection mechanisms used by \caratapp~\caratrefsonly.
Data collection by \caratapp is subject to the IRB process of UC
Berkeley\footnote{\url{http://cphs.berkeley.edu/}}.  For privacy
reasons, \caratapp{} does not collect any personally identifying
information about the user of the device.  \caratapp{} users are
informed about and provide consent to the data collected from their
devices.

The changes we made to \caratapp are to collect \dcvalue values of
packages in addition to the package names (\pvalue values) already
being collected.  Since \dcvalue values carry no additional
information about the user, our data collection technique does not
have any impact on the privacy of \caratapp users.

We have made our Carat dataset available for research
use\footnote{\url{http://se-sy.org/projects/malware/}}. In order to
protect the privacy of Carat users, we have made the following changes
to the shared dataset:

\squishlist
\item Compute a device pseudonym for the published data set by
  computing a randomized cryptographic hash (SHA-1) of \caratapp~ID
  using a fixed, secret salt.  This will prevent an adversary from
  correlating a device pseudonym with its \caratapp~ID or any of the
  other device identifiers that contributed to the computation of the
  \caratapp~ID.
\item Transform the package name (\pvalue) by computing its SHA-1
  hash. This is intended to hide the names of any unpublished package
  names that may have been present on a \caratapp~device of a
  developer or tester, while ensuring that the presence of a package
  in the dataset with a known package name can be verified.
  \squishend

\subsection{Malware Datasets}
\label{subsec:datasets:malware}

We used malware data from three different sources: the Malware
Genome dataset\footnote{\url{http://www.malgenomeproject.org/}}
provided by Zhou~et al.~\cite{zhou_dissecting_2012}, 
the Mobile Sandbox dataset\footnote{\url{http://mobilesandbox.org/}}
provided by Spreitzenbarth~\textit{et
  al.}~\cite{michael_spreitzenbarth_mobilesandbox:_2013}, and 
the McAfee dataset provided by
McAfee\footnote{\url{http://mcafee.com}}.  

The source of each malware dataset used their own unique set of
criteria to decide whether to include an Android package in the
dataset.  McAfee uses a proprietary classification technique.  Using
the Mobile Sandbox web interface, anyone can submit a package to Mobile Sandbox for
consideration.  Mobile Sandbox includes a package in their malware dataset if
any one of over 40 anti-virus tools they use flag the package as
malware. Malware Genome is a fixed set of known malware samples
collected during the period from August 2010 to October 2011~\cite{zhou_dissecting_2012}.

From each Android package (.apk file) in a dataset, we extract the package
identifier in the form of a \dcpv tuple. A malware
dataset is therefore a table of records, where each record is a
\dcpv tuple. Table~\ref{tab:malwaredatasets} 
summarizes the malware datasets.
%%%%%
\begin{table}[!htb]
  \centering{\small
    \begin{tabular}{|l|p{1.2cm}|r|p{1.2cm}|r|r|r|r|}
      \toprule
      Type                            & Mobile Sandbox         &
      McAfee     & Mobile Genome & Union \\
      \midrule
      unique dc           & 3,879      & 1,456  & 136         
%& 4,808      & 3,889     & 1,531       
& 4,809      \\
      unique \dcp       & 13,080     & 2,979  & 756 
%        & 15,014     & 13,196    & 3,473       
& 15,084     \\
      unique \dcpv     & 16,743     & 3,182  & 1,039
%       & 18,930     & 16,963    & 3,496       
& 19,094     \\
      unique .apk files     & 96,500     & 5,935  & 1,260
%       & 102,435    & 97,760    & 7,195       
& 103,695    \\
      \bottomrule
    \end{tabular}
    
    \caption{Summary of malware datasets.}
    \label{tab:malwaredatasets}
  }
\end{table}

\vspace{2em}
\section{Analysis of Infection Rates}
\label{sec:infectionrates}
The instrumentation of \caratapp{}---intended to be a battery saving app---is necessarily lightweight.  In
particular, we cannot perform detailed analysis of the applications on
the device to decide whether it is infected or not. Given the limited
information in the \caratapp{} dataset, we assessed infection rates by
examining if any of the malware packages (indicated by a malware
dataset) match any of the applications in the \caratapp{} dataset.

%\subsection{Matching with $<$dc$>$ and $<$dc,p,v$>$}
\subsection{Finding malware matches in \caratapp{} dataset}
\label{subsec:basicstatistics:matching}

We consider two types of ``match'': matching devcert only
(\dconly) and matching devcert, package name and version
(\dcpv). 
% We stress that we do not intend to propose the use of
% matching identifiers of packages with those of malware samples as the
% only technique for \textit{scanning} for malware on the device but as
% offline methods to estimate infection rates from historical
% application logs collected from the devices.

\noindent\textbf{\dconly}: One approach to identifying malware matches is to
deem a device in the \caratapp dataset as infected if it
has \textit{any} package signed with respect to a devcert found in a
malware dataset.  This way, when a \dcvalue associated with
a record in a malware dataset is seen in a \caratapp{} dataset record,
we flag that \caratapp{} record as infected. We proceed to compute the
number of unique bad devcerts ($N_C$), the number of unique packages
that correspond to bad devcerts ($N_P$), and the number of infected
devices as a whole ($N_{I:C}$). If the same devcert is used to sign
malware as well as benign packages, the \dconly approach may lead to
an over-estimate of infection rate.  However, it could serve as a way
to detect previously undetected malware, as Barrera~\textit{et
al.}~\cite{barrera_understanding_2012} point out.
\ifwww
% no algorithm in paper version
\else
Algorithm~\ref{alg:dconly} details the matching process.
\begin{algorithm}[!htb]
\centering
\begin{pseudocode}{Incidence\_dc\_only}{\mathcal{C}, \mathcal{M}}\label{alg:dconly}
\COMMENT{\caratapp dataset $\mathcal{C}$, Malware dataset $\mathcal{M}$}\\
\FOREACH \mbox{record} ~m \in \mathcal{M} \DO 
\IF \exists ~\mbox{record} ~c \in \mathcal{C} ~| ~c.dc = m.dc
\THEN c.status \GETS \mbox{infected}
\ELSE c.status \GETS \mbox{clean} \\ \\

\COMMENT{$S_C$: bad devcerts} \\
\COMMENT{$S_P$: \dcpv tuples w/ bad devcerts} \\
\COMMENT{$S_{I:C}$: infected devices} \\ \\

S_C \GETS \phi; S_P \GETS \phi; S_{I:C} \GETS \phi\\
\FOREACH \mbox{record} ~c \in \mathcal{C} \DO  
\IF ~c.status =  \mbox{infected}
\THEN 
\BEGIN
S_C \GETS S_C \cup \{c.dc\} \\
S_P \GETS S_P \cup \{c.dc|c.p|c.v\} \\
S_{I:C} \GETS S_{I:C}  \cup \{c.ID\}
\END \\
N_C \GETS |S_C|, N_P \GETS |S_P|, N_{I:C} \GETS |S_{I:C}| \\
\end{pseudocode}
\end{algorithm}
\fi

\noindent\textbf{\dcpv matching}: We can be more conservative by marking an entry in the \caratapp{} dataset as
infected if and only if every component of the reliable package
identifier (Section~\ref{subsec:background:android-packageid})
\dcpv of that entry
matches a record in the malware dataset.
This type of matching will underestimate the
infection rate, giving a lower bound.
\ifwww
% no algorithm in paper version
\else
The process is illustrated in Algorithm~\ref{alg:all}. 
\begin{algorithm}[!htb]
\centering
\begin{pseudocode}{Incidence\_dc+p+v}{\mathcal{C}, \mathcal{M}}\label{alg:all}
\COMMENT{\caratapp dataset $\mathcal{C}$, Malware dataset $\mathcal{M}$}\\
\FOREACH \mbox{record} ~m \in \mathcal{M} \DO 
\IF \exists ~\mbox{record} ~c \in \mathcal{C} ~| \\
(c.dc = m.dc ~\&\& \\ ~c.p = m.p ~\&\& \\ ~c.v = m.v ) 
\THEN c.status \GETS \mbox{infected}
\ELSE c.status \GETS \mbox{clean} \\ \\

\COMMENT{$S_{C,P}$: bad \dcpv tuples} \\
\COMMENT{$S_{I:C,P}$: infected devices} \\ \\

S_{C,P} \GETS \phi; S_{I:C,P,V} \GETS \phi\\
\FOREACH \mbox{record} ~c \in \mathcal{C} \DO  
\IF ~c.status =  \mbox{infected}
\THEN 
\BEGIN
S_{C,P} \GETS S_C \cup \{c.dc|c.p|c.v\} \\
S_{I:C,P,V} \GETS S_{I:C,P,V}  \cup \{c.ID\}
\END \\
N_{C,P,V} \GETS |S_{C,P,V}|, N_{I:C,P,V} \GETS |S_{I:C,P,V}| \\
\end{pseudocode}
\end{algorithm}
\fi
Table~\ref{tab:infection} summarizes the results of applying these two
approaches on the \caratapp{} dataset with the three malware datasets,
both separately and combined. We can make a number of observations.
First, there is a significant disparity in infection rates computed
using the two different approaches. Second, the number of infected
devices using \dcpv matching is largely disjoint for each malware
dataset, leading to the question of whether there is common agreement
as to what constitutes malware.  In subsections \ref{subsec:dc-vs-dcpv}
and \ref{subsec:whatismalawre}, we examine these issues
in more detail.  No \dcpv tuples from the \caratapp data matched the two-year-old
 Malware Genome dataset.
This suggests that in the long run, malware on Android devices is detected and removed.
% The infection rate with all malware
% datasets is 0.7\%. Table~\ref{tab:frequent} show the top of frequent
% malware, and it also shows the disagreement about malware.

\begin{table*}[!htb]
  \centering 
  {\small    
    \begin{tabular}{|l|r r|r r|r r|r r|}
      \toprule
      \hspace{5cm}\hfill Malware dataset& Mobile &            & McAfee     &         & Malware &  & Union of&  \\
      Type & Sandbox & \%          & & \%        & Genome & \% & all sets& \% \\
      \midrule
      $N_C$: \# of \dconly matches (bad devcerts)                         & 337 & 0.81 & 343 & 0.82 & 8 & 0.019 & 534 & 1.3 \\
      %337            & 343              & 8            & 534         \\
      $N_P$: \# of packages (\dcpv) matching malware $<$dc$>$             & 10,030 & 5.2 & 10,062 & 5.2 & 6137 & 3.2 & 11,510 & 6 \\
      %10,030         & 10,062           & 6,137        & 11,510      \\
      $N_{C,P,V}$: \# of packages matching malware \dcpv                   & 100 & 0.15 & 62 & 0.096 & 0 & 0 & 155 & 0.24 \\
      %100             & 62              & 0            & 155         \\
      $N_{I:C}$: \# of infected devices (\dconly match)                   & 18,719 & 34 & 16,416 & 30 & 9240 & 17 & 20,182 & 37 \\ 
      %18,719 (34\%)   & 16,416 (30\%)   & 9240(16.7\%) & 20,182 (41\%)\\
      $N_{I:C,P,V}$: \# of infected devices (\dcpv match)                  & 154 & 0.28 & 144 & 0.26 & 0 & 0 & 285 & 0.52 \\
%      154 (0.28\%)    & 144 (0.26\%)    & 0(0\%)       & 285  (0.51\%) \\
      \bottomrule
    \end{tabular}
  }
  \caption{Incidence of infection in the \caratapp{} dataset.
  %MB: Mobile Sandbox, MG: Malware Genome.
  \label{tab:infection}}
\end{table*}

\subsection{Disparity in $<$dc$>$  vs. $<$dc,p,v$>$ Matching}
\label{subsec:dc-vs-dcpv}

\begin{figure*}[!htb]
  \centering
  \begin{minipage}{0.35\textwidth}
    \includegraphics[width=\textwidth]{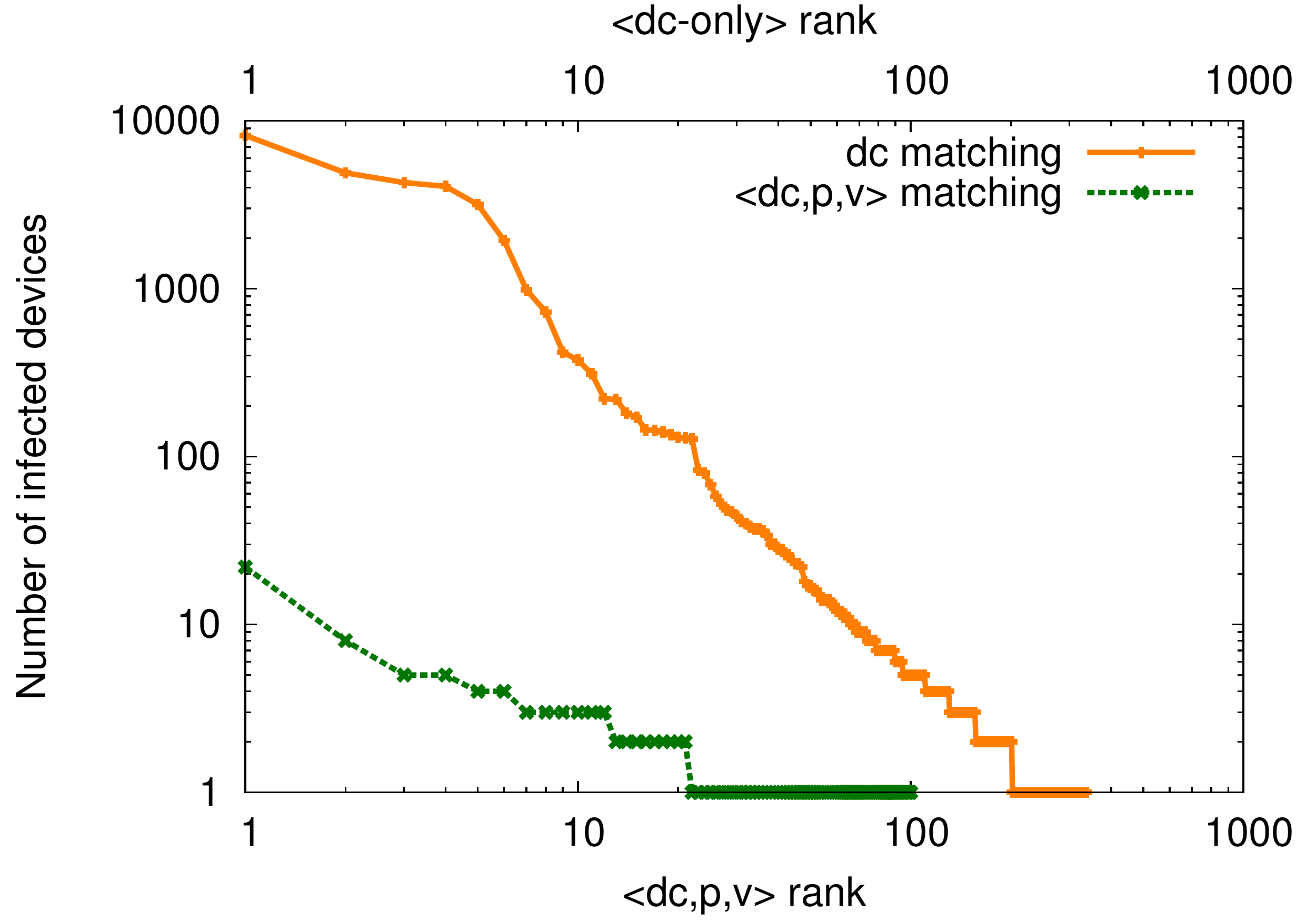}
    \subcaption{Mobile Sandbox dataset}
    \label{fig:package-device-MB}
  \end{minipage}  
%%%
  \begin{minipage}{0.35\textwidth}
    \includegraphics[width=\textwidth]{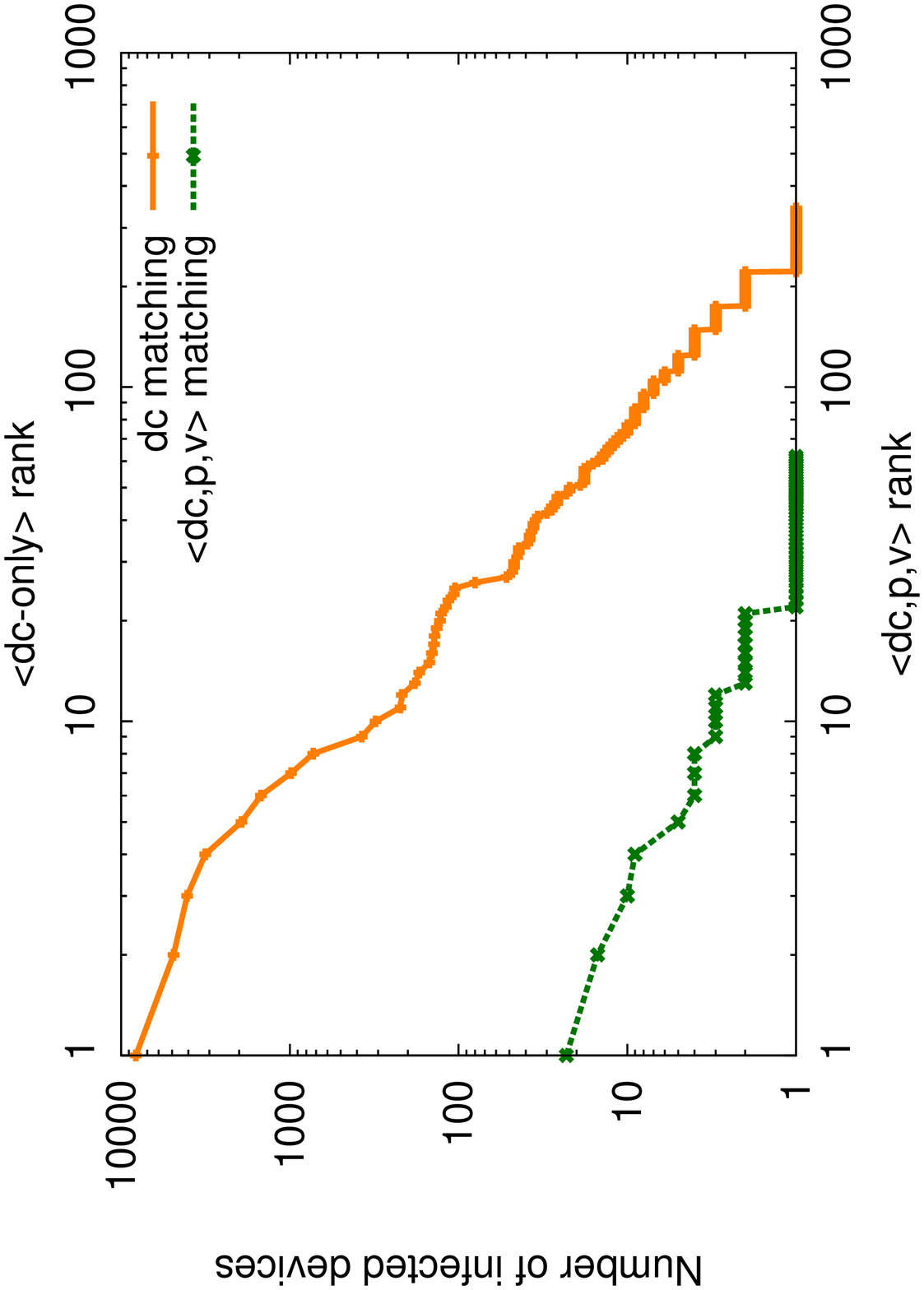}
    \subcaption{ McAfee dataset}
    \label{fig:package-device-MC}
  \end{minipage}  
  \caption{The number of infected devices based on \dconly matching is orders of magnitude larger than with \dcpv matching.}
  \label{fig:infection}
\end{figure*}

Figure \ref{fig:infection} shows the distribution of infected devices
using \dconly and \dcpv matching.
We now discuss two reasons for the
discrepancy between \dconly matching and \dcpv matching.
\begin{table*}[htb]
%\begin{sidewaystable}[htbp] %In case table need to span over the length size
  \centering
   {\small
   \begin{adjustwidth}{-0.2cm}{}
   \begin{minipage}{18.25cm}
   \begin{tabularx}{\linewidth}{@{}|l|l|p{0.8cm}|p{0.8cm}|p{1.5cm}|p{2.2cm}|@{}}
      \toprule
      
      Package name 
      & Package hash MD5 \& SHA1 
      & \#inf\footnote{Number
        of devices infected by this package} 
      & \#det\footnote{Number
        of anti-malware tools flagging this package as malware according to \url{http://mobilesandbox.org}}
      & Description 
      %& P\footnote{Potentially unwanted program (PUP)?} \\
      & Source \\
      \cmidrule{1-6}

      \multirow{2}{*}{it.evilsocket.dsploit} & 
      7dedeca3c23462202f86e52dcc004231 &       
      \multirow{2}{*}{23}  & 
      \multirow{2}{*}{11} & 
      \multirow{2}{*}{Monitoring} & 
      \multirow{2}{*}{McAfee} \\
      & 07d522ab602352ab27523f92f94bd2ee9b10e40e & & & & \\

      \multirow{2}{*}{com.noshufou.android.su} & 
      1e74b778765d2e49caaa06772c68edbc & 
      \multirow{2}{*}{23} & 
      \multirow{2}{*}{17} & 
      \multirow{2}{*}{Rooting} &
      \multirow{2}{*}{McAfee} \\
      & 600621d4c04fb3a699c292cd7aec21676510351d & & & & \\

      \multirow{2}{*}{ty.com.android.SmsService} &
      af15196deb350db09eafe3e1cb751f2e & 
      \multirow{2}{*}{22} & 
      \multirow{2}{*}{4} & 
      \multirow{2}{*}{Trojan} & 
      \multirow{2}{*}{Mobile Sandbox} \\
      & 7658d936fd559c3dbe0d30ab669269c5d0e7e512 & & & & \\

      \multirow{2}{*}{com.mixzing.basic} &
      30cf001554ef0c10bc0012e3a75950ba &
      \multirow{2}{*}{15} & 
      \multirow{2}{*}{15} & 
      \multirow{2}{*}{Adware} & 
      \multirow{2}{*}{McAfee} \\
      & 6dc4477f570d77fdb4adf5db3ee3347a2e9a4b11 & & & & \\

      \multirow{2}{*}{pl.thalion.mobile.battery} & 
      2e0a763029636ee3555c59f92bd8bd92 &
      \multirow{2}{*}{10} & 
      \multirow{2}{*}{11} & 
      \multirow{2}{*}{Adware} & 
      \multirow{2}{*}{McAfee} \\
      & 4cf39278dba8453f1f58c28d08a8478bb2f18a30 & & & & \\

      \multirow{2}{*}{com.bslapps1.gbc} &
      e911ba9d519c36eb02b19268ee67de35 &
      \multirow{2}{*}{9} & 
      \multirow{2}{*}{7} & 
      \multirow{2}{*}{Adware} & 
      \multirow{2}{*}{McAfee} \\
      & f00ab5a4d720fc1489530354cb9fd6d11242a77b & & & & \\

      \multirow{2}{*}{com.android.antidroidtheft} &
      2d6130aaa0caa1a170fb0ed1c0d687c7 &
      \multirow{2}{*}{8} & 
      \multirow{2}{*}{3} & 
      \multirow{2}{*}{Monitoring} & 
      \multirow{2}{*}{Mobile Sandbox} \\
      & fcfa52c70c29f0b6712cb02ba31c053fbcf102e4 & & & & \\

      \multirow{2}{*}{com.androidlab.gpsfix} &
      9f6e1d4fad04d866f1635b20b9170368 &
      \multirow{2}{*}{7} & 
      \multirow{2}{*}{9} & 
      \multirow{2}{*}{Adware} & 
      \multirow{2}{*}{McAfee} \\
      & e1c1661a4278d572adbf7439f30f1dcc1f0d5ea5 & & & & \\

      \multirow{2}{*}{com.adhapps.QesasElanbiaa} &
      3a818a3a8c8da5bebbefdc447f1f782f &
      \multirow{2}{*}{7} & 
      \multirow{2}{*}{15} & 
      \multirow{2}{*}{Adware} & 
      \multirow{2}{*}{McAfee} \\
      & 7b8d16c362e8ac19ceed6ec0e43f837ee811ac7a & & & & \\      

      \multirow{2}{*}{download.youtube.downloader.pro7} &
      6bad5fa9b87d0a5d7e81994d8e6e4d38 &
      \multirow{2}{*}{5} & 
      \multirow{2}{*}{6} & 
      \multirow{2}{*}{Adware} & 
      \multirow{2}{*}{Mobile Sandbox}  \\
      & 074385ac4938cadc1f93f4a92b811786d2f01ac6 & & & & \\      

      \multirow{2}{*}{com.android.settings.mt} &
      fa037c0c6dcfe0963d9e243ee6989dc1 &
      \multirow{2}{*}{5} & 
      \multirow{2}{*}{4} & 
      \multirow{2}{*}{Monitoring} & 
      \multirow{2}{*}{McAfee} \\
      & c1c72cd10f3714c2fb4b1ca281b5f33192d2d872 & & & & \\           
      
      \bottomrule
    \end{tabularx}
    \end{minipage}
    \end{adjustwidth}    
  }
  \caption{Most frequent malware. \label{tab:frequent}}
\end{table*}
%\end{sidewaystable} %% In case table span the length size

\textbf{Reuse of devcerts:}
The first reason is that some devcerts are used to sign both malware and
clean packages. Consider the case of the popular Brightest Flashlight
Free application: v17 was flagged as malware by a number of
anti--malware vendors, presumably because an ad library it was using
was malicious.  Subsequent versions are not flagged as malware.  All
versions, however, are signed with respect to the same
devcert\footnote{\url{https://androidobservatory.org/cert/27DDACF8860D2857AFC62638C2E5944EA15172D2}},
which appeared in $2210$ devices in our \caratapp dataset, but never
with v17.  The Android ``same-origin'' policy that allows applications
signed using the same developer key to share data among themselves
discourages (even legitimate) developers from changing their devcerts
even if an earlier version of their package was flagged as malware.

A malware developer may also have non-malware packages, and it is
known that malware developers are actively seeking to purchase
verified developer accounts~\cite{brian_krebs_mobile_2013_wurl}.

\begin{figure*}[!htb]
  \centering
  \begin{minipage}{0.35\textwidth}
    \includegraphics[width=\textwidth]{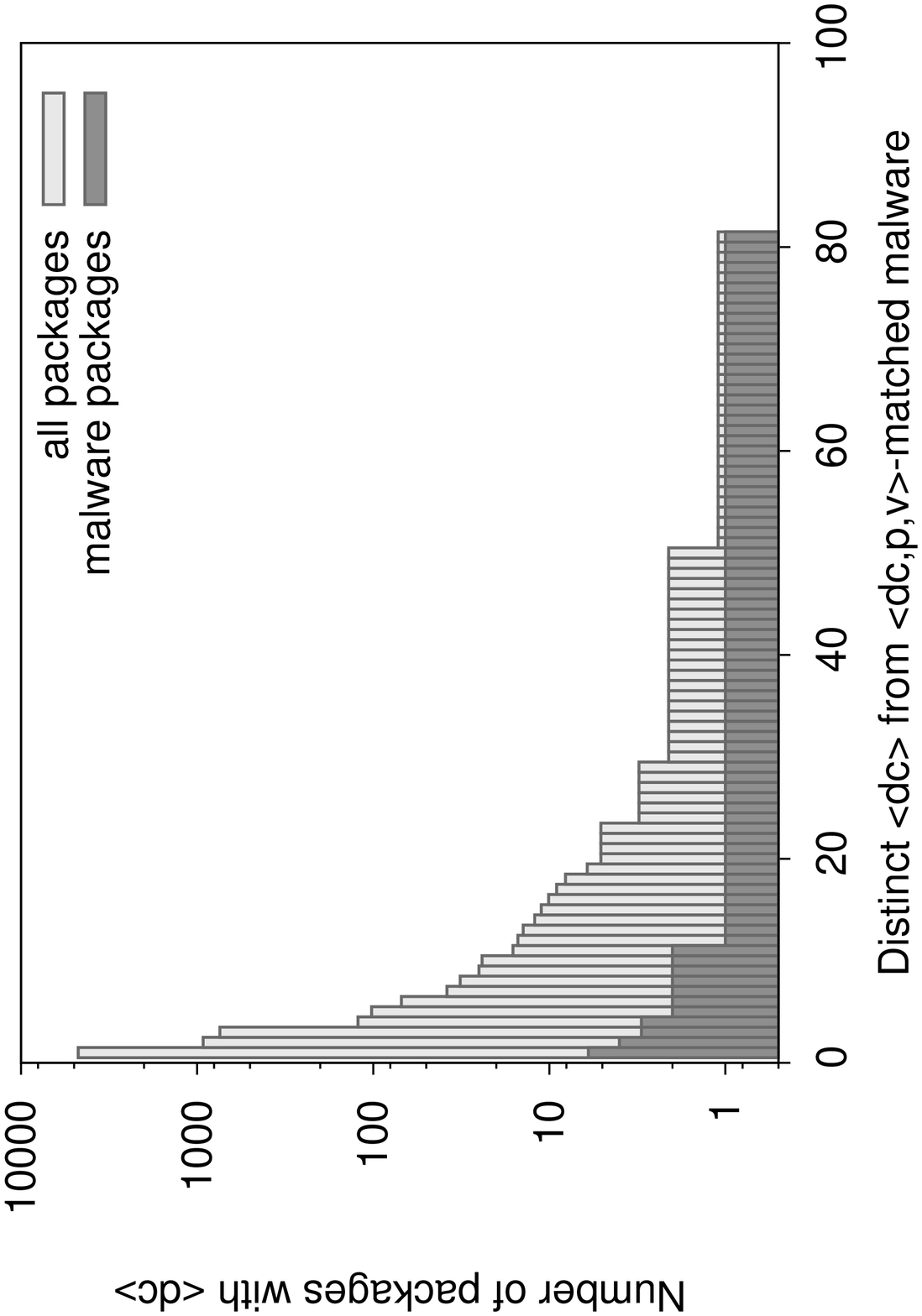}
    \subcaption{Mobile Sandbox dataset}
    \label{fig:dc-package-MB}
  \end{minipage}  
%%%
  \begin{minipage}{0.35\textwidth}
    \includegraphics[width=\textwidth]{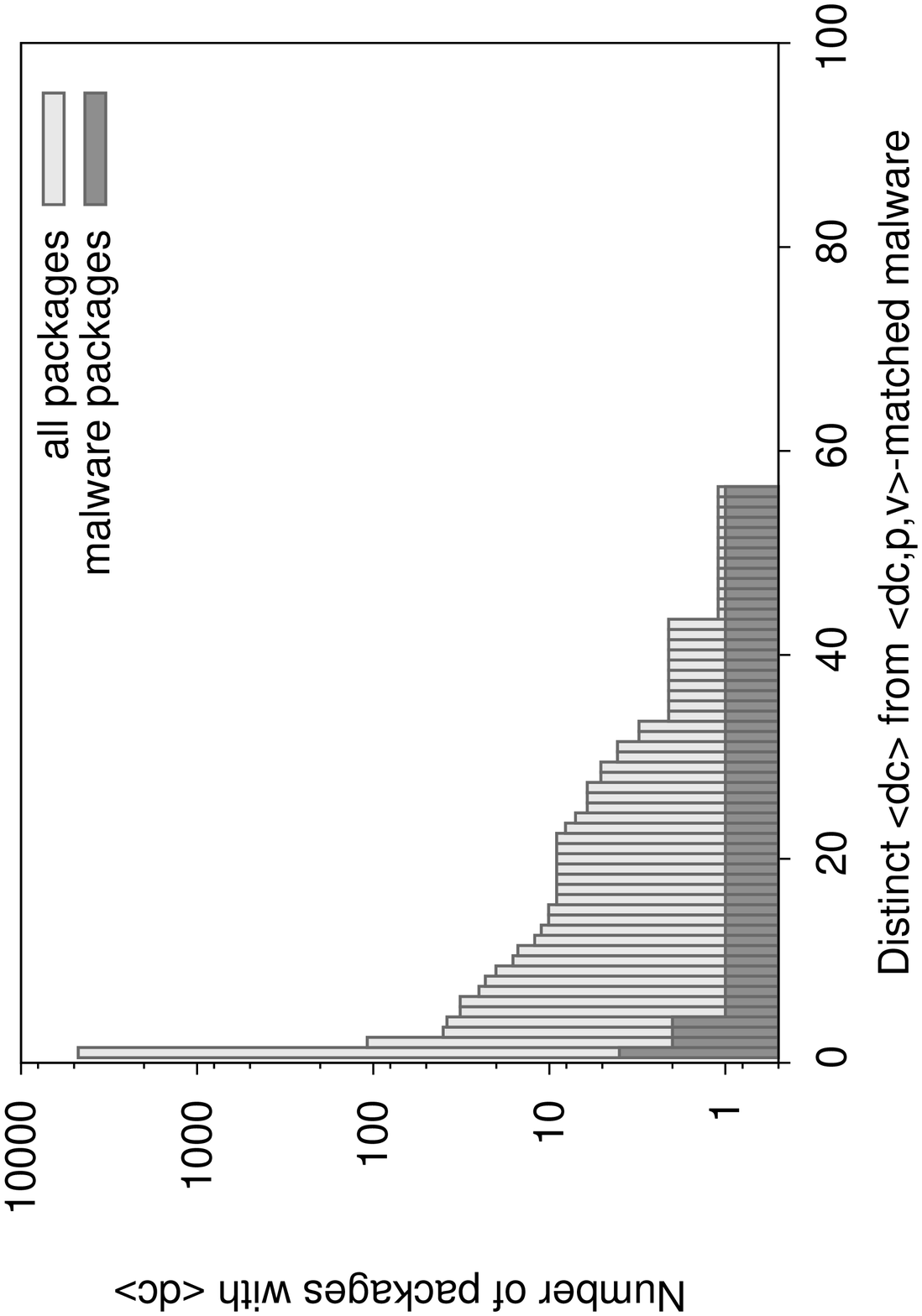}
    \subcaption{ McAfee dataset}
    \label{fig:dc-package-MC}
  \end{minipage}  

  \caption{Packages signed with respect to a potentially malicious <dc>.}  
  \label{fig:dc-packages}
\end{figure*}

\textbf{Widely available signing keys:}
The second reason is that some signing keys are widely available
(either because they were leaked or because they were test keys).
Surprisingly, some legitimate developers use these widely available
signing keys to sign their packages.  One particular
devcert\footnote{\url{https://androidobservatory.org/cert/61ED377E85D386A8DFEE6B864BD85B0BFAA5AF81}}
is a widely available ``test key''. $544$ packages in our malware
datasets were signed with it. However, $1948$ innocuous packages in
our
\caratapp{} dataset (several of them used to be in Google Play) were
also signed with the same key.
In the \caratapp{} dataset, $8188$
devices had at least one package signed with it but only $11$ devices
had a package that matched the full \dcpv of a known malware package.

In all of these cases, the same key is used to sign both malware and
non-malware. Consequently the \dconly method of identifying
applications results in an overestimate of malware infection
rate. While it provides an upper bound of infection rate, the estimate
is too high to be useful, marking more than $16\%$ of devices as infected
for all datasets, and over $30\%$ for Mobile Sandbox and McAfee.
Therefore, in the rest of this paper, we
use only \dcpv matching.

Figure \ref{fig:dc-packages} shows more details of how devcerts
associated with malware are reused. Unsurprisingly, most malware
devcerts are used to sign only one package each.  However out of a total of
155 devcerts associated with malware (in both Mobile Sandbox and McAfee
malware datasets taken together), there are still 13 devcerts signing
more than one malware package, and 70 devcerts signing more than one
package in general.

Interestingly, we found that \texttt{com.android.settings.mt} signed
by a
devcert\footnote{\url{https://androidobservatory.org/cert/9CA5170F381919DFE0446FCDAB18B19A143B3163}}
controlled by Samsung was flagged as malware by multiple anti-malware
vendors\footnote{\url{http://bit.ly/IFDMyl}}. (The same key is widely
used by Samsung to sign their packages including some available on
Google Play like
\texttt{com.sec.yosemite.phone}, and
\texttt{com.airwatch.admin.samsung}).  Samsung has since updated the
package, which is no longer flagged as malware, but continues to use
the \textit{same} the version code (1) for all versions.
Consequently, \dcpv matching resulted in devices containing all
variants of this package (7845 devices in the \caratapp{} dataset)
being labeled as infected. We were therefore forced to discount this
package from our analysis because we have no way to distinguish
between its malware and non-malware variants.  We maintain that \dcpv
matching is still a reasonable approach for identifying malware
because ordinary developers are required to update the version code when
they release new
versions\footnote{\url{http://developer.android.com/tools/publishing/versioning.html}}.
%%%%%
\vspace{2em}
\subsection{Disagreement about What is Malware}
\label{subsec:whatismalawre}
\begin{figure}[!htb]
  \centering
   \includegraphics[width=0.6\columnwidth]{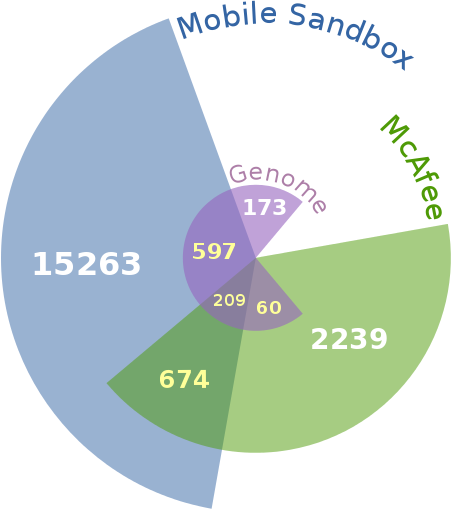}
  \caption{Sizes of the three malware datasets and the extent of
    overlaps among them.}
  \label{fig:malware}
\end{figure}

%%% 
Figure~\ref{fig:malware} shows the distribution of malware packages in
the three datasets we used.  As can be seen, a significant fraction
of each dataset contains malware samples included only in that set,
leading to the question whether there is any common agreement about
what constitutes malware.  This issue is illustrated even more
dramatically in Table~\ref{tab:infection}, which shows the number of
devices labeled as infected according to each individual malware
dataset. There were 154 and 144 devices labeled as infected according
to the Mobile Sandbox and McAfee datasets, respectively, but only 13
devices were common to these sets.
%%%%%%%
%%%%%%%
Table~\ref{tab:frequent} lists the most frequent malware
packages---those that were detected in five or more devices---in our \caratapp{}
dataset. Each package was scanned by over 40 different anti-malware
tools; \emph{none} of the malware packages that match our \caratapp{}
dataset is flagged as malware by a majority of anti-malware tools.

All of these observations confirm that there is no wide agreement
among anti-malware tools about what constitutes malware.  Given the
extent of disagreement, we conclude that it is more appropriate to
report infection rates separately for each malware dataset
(\MobSandinfectionrate for Mobile Sandbox and \McAfeeinfectionrate for
McAfee) rather than combined (\infectionrate).

One reason for the disagreement is that some
packages are considered ``potentially unwanted programs'' (PUPs) by
some anti-malware tools, but
not outright malware.  For example, Superuser (\texttt{com.noshufou.android.su}),
a popular rooting application, cannot be considered
malicious if it is installed intentionally by the user but is
certainly dangerous if installed without the user's knowledge.
Similarly, a WiFi-cracking tool is dangerous but not malicious towards
the user. Some anti-malware tools consider applications
containing intrusive adware libraries as malware while others do
not. Some vendors eschew the term ``malware'' altogether and resort to
identifying threats in different categories separately~\cite{Lookout13}.
%The last column of
%Table~\ref{tab:frequent} lists the packages we consider PUPs. 

On the other hand, the disagreement among anti-malware tools is
puzzling because there is evidence that anti-malware vendors use
systems that monitor how other vendors rate a package and incorporate
this information into their own rating.  Sometimes, this can lead to
mistakes being propagated widely.  We encountered one such example in
our dataset in the form of the package
\texttt{com.android.vending.sectool.v1} signed with respect to a
legitimate devcert, controlled by
Google\footnote{\url{https://www.androidobservatory.org/cert/24BB24C05E47E0AEFA68A58A766179D9B613A600}}.
It corresponds to a malware removal tool published by Google almost
three years
ago\footnote{\url{http://googlemobile.blogspot.com/2011/03/update-on-android-market-security.html}}.
Shortly thereafter, a malware package with the same name
emerged~\cite{AMTBackdoor11}\footnote{\url{http://bit.ly/IDjR3W}}.
Presumably some anti-malware vendor mistakenly labeled \text{both}
packages as malware which consequently led the some of the rest to
follow suit.  The legitimate Google-signed package is flagged as
malware by no less than 13 anti-malware tools to this
day\footnote{\url{http://bit.ly/IDjWos}}.
% In the
% rest of the discussion, we choose to include PUPs because a number of
% anti-malware tools block each PUP.  in Table~\ref{tab:frequent}.  For
% example, the ACM mail gateway, which handles \texttt{@acm.org} e-mail
% aliases, blocks one of those packages, if they are sent as attachments
% in an e-mail message.  Furthermore, we speculate that enterprise IT
% adminstrators will block PUPs like rooting applications as unwanted.
%%%%
%%%%%
\subsection{Geographic Regions}
\label{subsec:predictions:language}
Mobile malware infection rates in China and Russia are reported to be
higher than in the West~\cite{Lookout12}.  Since we do not have
demographic information about \caratapp users, we do not know their
location and thus cannot verify this claim directly.  However, the
translated name (transname) of a package can sometimes indicate the
local language of the device.  The length of the string is too short
for robust automated language detection. 
%herefore we do not report infection
%ate grouped by language.
%% Most devices had only packages with English names.  These were
%% classified as English.  Among the remaining devices, five or more
%% packages were named in a second language.  These were assigned the
%% second language (there was no device with packages named in more than
%% two languages).  The results are in Table ~\ref{tab:lang}.
However, we can, based on the presence of operator-specific (e.g.,
\texttt{com.vzw.hss.myverizon}, ``My Verizon Mobile'') and
currency-specific (e.g., \texttt{com.fdj.euro}, \newline
``Euro Millions'') package names, estimate the number of
infected devices in certain regions like the US, Europe, and Japan.
\ifwww
Consequently, we can conjecture that the infection rate in USA
is likely to be more than 0.02\% based on the fact that 13 infected
devices (as per McAfee malware dataset) have USA-specific packages.
\else
(See Table~\ref{tbl:lang}.)
This suggests that the infection rate in USA is
likely to be more than 0.02\% (based on the lower estimate of 13
infected devices in USA in Table~\ref{tbl:lang}).
\fi
%% at least $33$ of English language devices are from the United States,
%% $7$ devices are in Europe and $6$ from Japan.
%%%
\ifwww
\else
\begin{table}[!htb]
  \centering
  \begin{tabular}{|l|c|c|}
      \hline
      \multirow{2}{*}{Location} & \multicolumn{2}{c|}{\# infected devices}   \\
      \cline{2-3}
                                &  w.r.t Mobile Sandbox       & w.r.t McAfee   \\
      \hline
      USA     & 22              & 13                \\
      Europe  & 2                & 1               \\
      Japan   & 5                & 3                 \\
      \hline
    \end{tabular}
  \caption{Lower bounds for infected devices in some localities.}
  \label{tbl:lang}
\end{table}
\fi

\section{Detecting Potential Infection}
\label{sec:predictions}
%%%!TEX root = submission.tex
Mobile device owners, enterprise IT administrators, and vendors of
anti-malware tools would all benefit from techniques that can provide
early warnings about the susceptibility to malware infection. In this
section, we report on the in-depth analyses of our data and look at factors
that have a strong influence on the risk of malware infection.

%analyze factors that influence the risk of infection and assess how these could be used to 
\ifwww
\else
\subsection{Energy Consumption}
\label{subsec:predictions:energy}
\begin{table}[!htb]
\centering
\small{
\begin{tabular}{|l|c c|c c|}
\hline
& \multicolumn{2}{c|}{\textbf{Mobile Sandbox}} & \multicolumn{2}{c|}{\textbf{McAfee}} \\
\hline
\textit{Statistic} & \textit{Infected} & \textit{Clean} & \textit{Infected} & \textit{Clean} \\
\hline
\multicolumn{5}{c}{\textbf{All Models and OS Versions}} \\
\hline
Mean & 7.28 & 9.84 & 	 8.37 & 9.84 \\ 
Median & 6.39 & 8.14 & 	 8.08 & 8.13 \\ 
Difference (\%) & \multicolumn{2}{c|}{26} & \multicolumn{2}{c|}{14.9} \\ 
Wilcoxon & \multicolumn{2}{c|}{p < 0.01, Z=-5.26} & 	 \multicolumn{2}{c|}{p=0.177, Z=-1.35} \\ 
Pearson corr & \multicolumn{4}{c|}{-0.0471}\\ 
\hline
\multicolumn{5}{c}{With 95th percentile high outlier removal} \\ 
\hline
Mean & 6.56 & 8.36 & 	 7.88 & 8.35 \\ 
Median & 6.06 & 7.94 & 	 7.74 & 7.94 \\ 
Difference (\%)  & \multicolumn{2}{c|}{21.5} & 	 \multicolumn{2}{c|}{5.63} \\ 
Wilcoxon & \multicolumn{2}{c|}{p < 0.01, Z=-5.68} & 	 \multicolumn{2}{c|}{p=0.152, Z=-1.43} \\ 
Pearson corr & \multicolumn{2}{c|}{0.0202}& 	 \multicolumn{2}{c|}{0.0202}\\ 
\hline
\multicolumn{5}{c}{\textbf{Matching Models and OS Versions}} \\
\hline
Mean & 7.28 & 9.03 & 	 8.37 & 9.52 \\ 
Median & 6.39 & 7.68 & 	 8.08 & 8.08 \\ 
Difference (\%)  & \multicolumn{2}{c|}{19.3} & 	 \multicolumn{2}{c|}{12.1} \\ 
Wilcoxon & \multicolumn{2}{c|}{p= < 0.01, Z=-3.89} & 	 \multicolumn{2}{c|}{p=0.261, Z=-1.12} \\ 
Pearson corr & \multicolumn{2}{c|}{-0.0481}& 	 \multicolumn{2}{c|}{-0.0477}\\ 
\hline
\multicolumn{5}{c}{With 95th percentile high outlier removal} \\ 
\hline
Mean & 6.56 & 7.88 & 	 7.88 & 8.27 \\ 
Median & 6.06 & 7.5 & 	 7.74 & 7.9 \\ 
Difference (\%)  & \multicolumn{2}{c|}{16.8} & 	 \multicolumn{2}{c|}{4.67} \\ 
Wilcoxon & \multicolumn{2}{c|}{p < 0.01, Z=-4.22} & 	 \multicolumn{2}{c|}{p=0.23, Z=-1.2} \\ 
Pearson corr & \multicolumn{2}{c|}{-0.0401}& 	 \multicolumn{2}{c|}{-0.0161}\\
\hline
\end{tabular}
}
\caption{Statistics on battery life.\label{tbl:battresults}}
\end{table}
\caratapp{} provides accurate estimates of the average battery life of
a device as long as a sufficient number of samples submitted
by the device is available. We use these estimates to compare
differences in battery life between infected and clean (uninfected)
devices. There was sufficient data for \caratapp{} battery life estimates on $112$ infected devices (out of $154$)
for Mobile Sandbox, and $114$ infected devices (out of $144$) for McAfee. To ensure
software and hardware variations did not cause bias in the analysis,
only those clean devices that have the same model and OS version as at
least one of the infected devices were included in the analysis,
resulting in a sample of $9,626$ clean devices for Mobile Sandbox and $12,766$
for McAfee.

We removed outliers beyond the 95th percentile of battery life.
481 outliers were removed from 
clean and 6 from infected data for Mobile Sandbox. For McAfee,
we removed 638 from clean and 6 from infected battery life data.

The mean battery life after outlier removal for the infected devices
was $6.56$ hours (median $6.06$), and $7.88$ hours for the clean devices
(median $7.5$) for Mobile Sandbox. A Wilcoxon rank sum test indicated
statistical significance in the lifetime distributions (Z = -4.2215, p
< 0.01).

%Pearson correlation coefficient of installed applications vs battery life for all data: -0.048129
%With 95th percentile outlier removal:
%Pearson correlation coefficient of installed applications vs battery life for all data: -0.040077
%
For McAfee, the mean battery life was $7.88$ hours for infected (median
$7.74$) and $8.27$ hours for clean (median $7.9$). The difference was
marginally significant (Z = -1.199, p = 0.23).

Without outlier removal, and without limiting the OS and device model
of clean devices, the same relationship holds; the difference in battery
life between infected and clean is significant for Mobile Sandbox and
marginally significant for McAfee (see Table~\ref{tbl:battresults} for
details).

%Pearson correlation coefficient of installed applications vs battery life for all data: -0.047674
%With 95th percentile outlier removal:
%Pearson correlation coefficient of installed applications vs battery life for all data: -0.016100

%Comparison between the lifetimes 
%indicated a marginal significance in the lifetime distributions 
%As the lifetime distributions are heavily skew

%The standard deviation was $2.1358$ hours.
%The number of points was $887$ and $14$ respectively.
%The difference in battery life between clean and infected devices
% (see Figure~\ref{fig:energy}) 
%is statistically significant according to an independent samples t-test with Welch correction $(t =
%3.7709, df = 35, p < .0004)$.

% \begin{figure}[htbp]
%   \centering
%   \includegraphics[width=0.475\textwidth]{figures/energy_comparison_boxplot.png}
%   \caption{Comparison of battery life between clean and infected devices. The y-axis is average battery lifetime in hours.}
%   \label{fig:energy}
% \end{figure}

The reduced lifetime for the infected devices could be caused by a
difference in the number of applications that are used on the
devices. To demonstrate that this is not the case, we next investigate
the relationship between battery consumption and application
usage. Pearson's correlation coefficient between the two was $-0.0401$
for Mobile Sandbox and $-0.0161$ for McAfee, suggesting they are
uncorrelated. A small positive value was observed for all models and
OSes with outlier removal, however, the magnitude indicates no
correlation.  Furthermore, there are examples of clean devices with
more than 200 apps and a higher than average battery life.  An
infected device with over 300 applications installed had 16 hours of
battery life, over 6 hours more than the mean for clean devices 
in all the cases. The number of applications does not
seem to correlate with high energy use.

This is also illustrated in Figure~\ref{fig:pkg-installed}, which
shows the average battery life for the two device groups.  The
detailed statistics are shown in Table~\ref{tbl:battresults}.

%%%
\begin{figure*}[!htb]
  \centering
  \begin{minipage}{0.42\textwidth}
    \centering
    \includegraphics[width=0.8\textwidth]{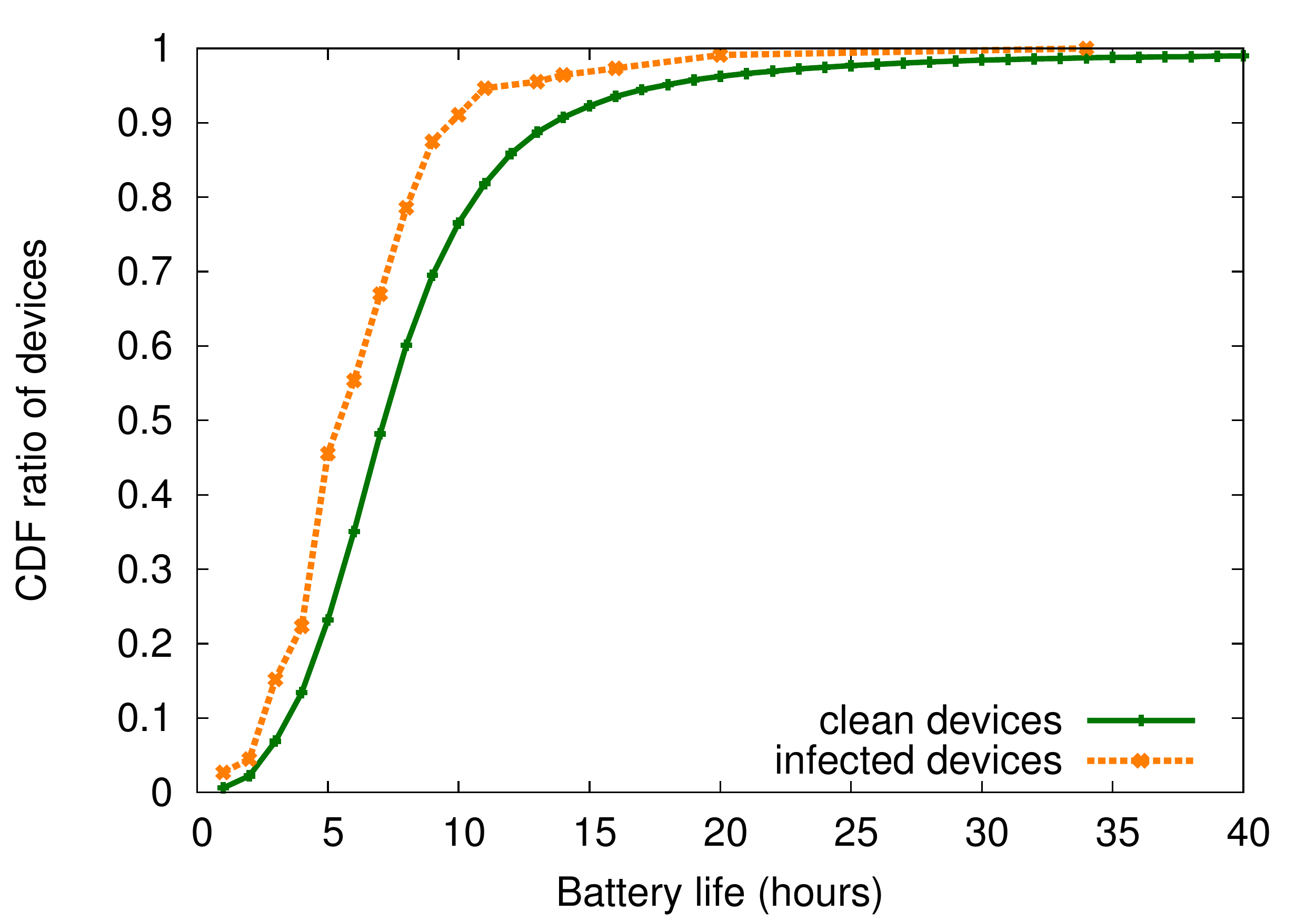}
    \subcaption{Mobile Sandbox dataset}
    \label{fig:package-device-dc-package-MB}
  \end{minipage}  
%%%
  \begin{minipage}{0.42\textwidth}
    \centering
    \includegraphics[width=0.8\textwidth]{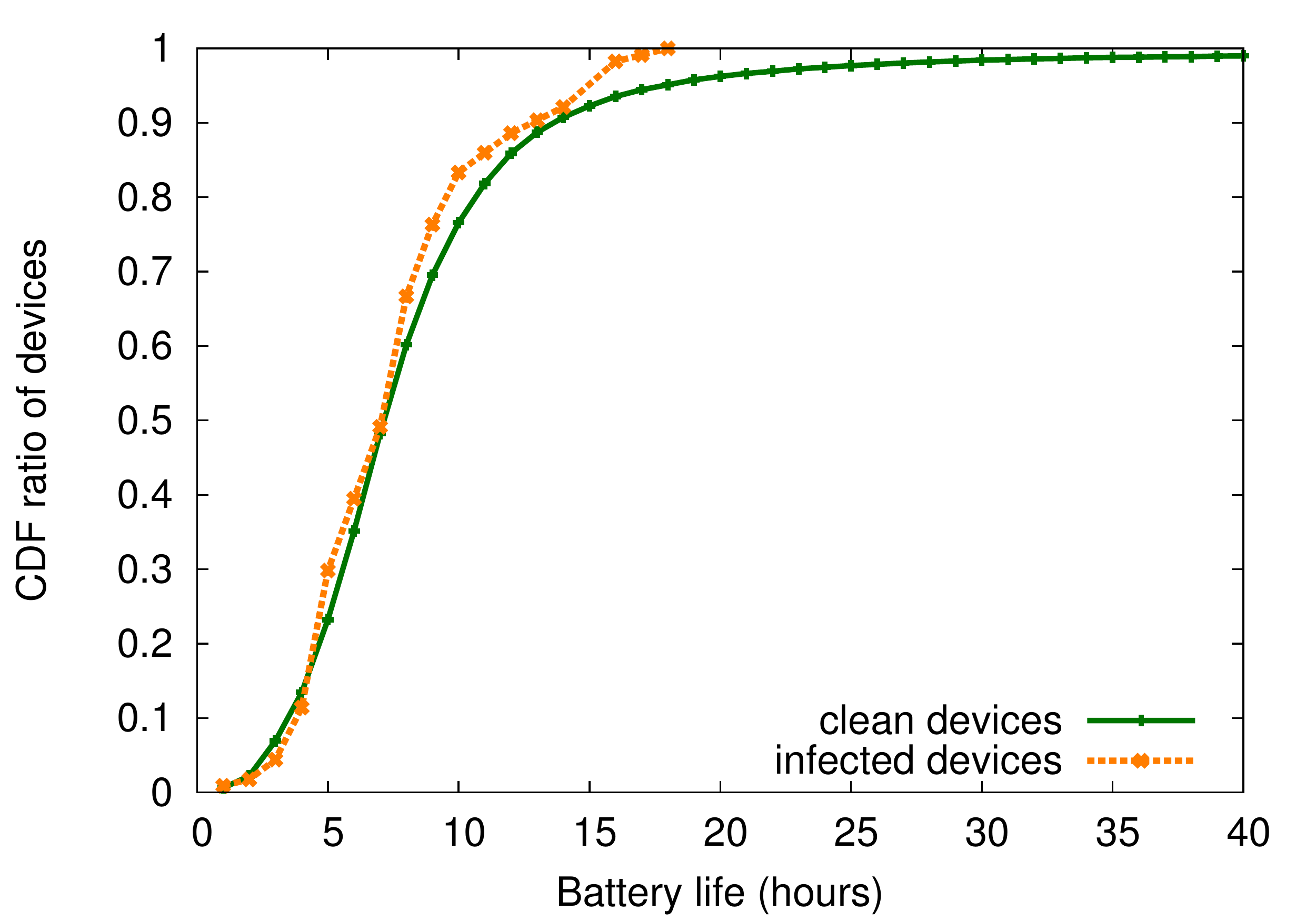}
    \subcaption{ McAfee dataset}
    \label{fig:package-device-dc-package-MC}
  \end{minipage}  
  \caption{Average battery life. \label{fig:energy}}
\end{figure*}
%%%%

%%%
\begin{figure*}[!htb]
  \centering
  \begin{minipage}{0.42\textwidth}
    \centering
    \includegraphics[width=0.8\textwidth]{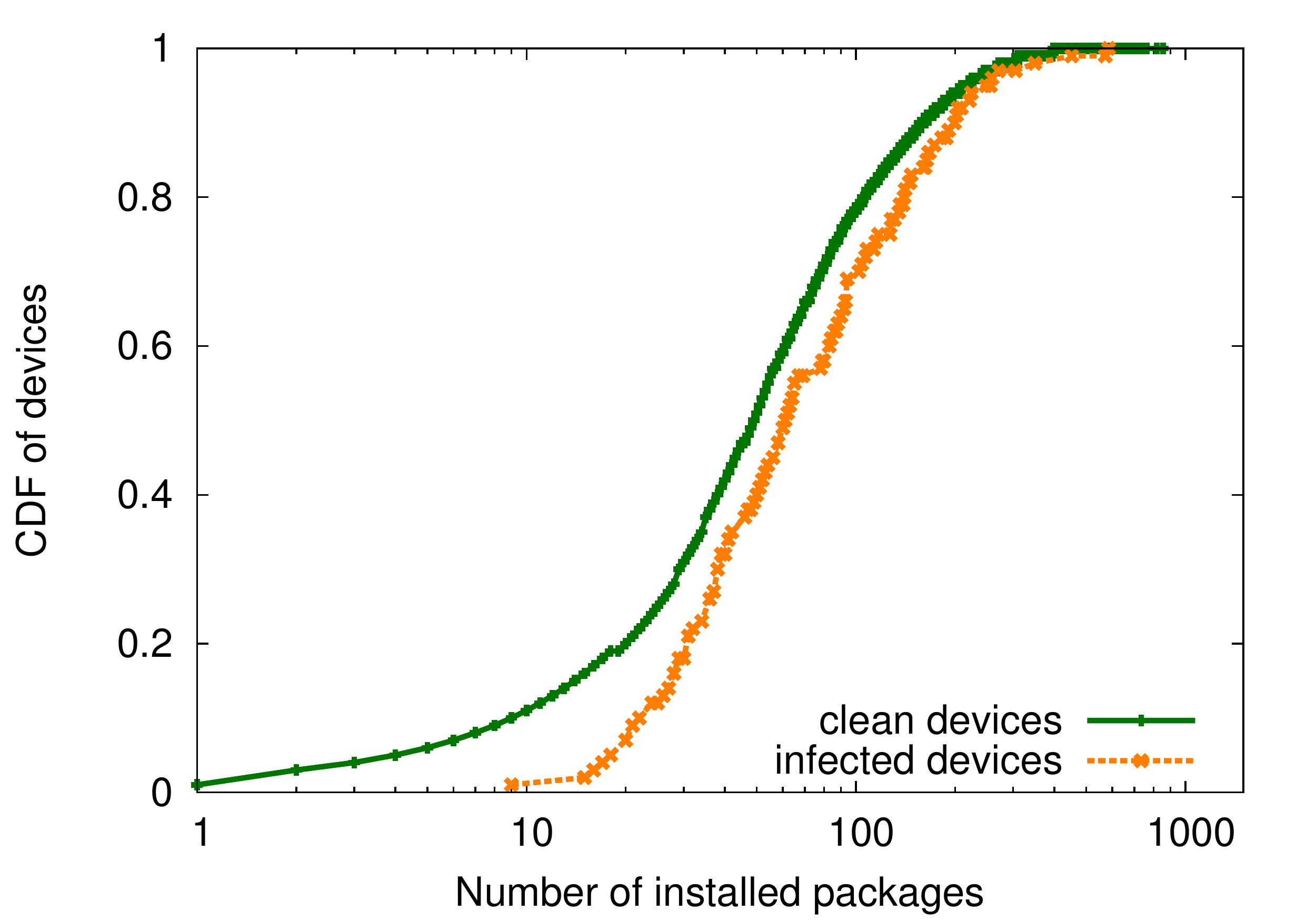}
    \subcaption{Mobile Sandbox dataset}
    \label{fig:package-device-dc-package-MB}
  \end{minipage}  
%%%
  \begin{minipage}{0.42\textwidth}
    \centering
    \includegraphics[width=0.8\textwidth]{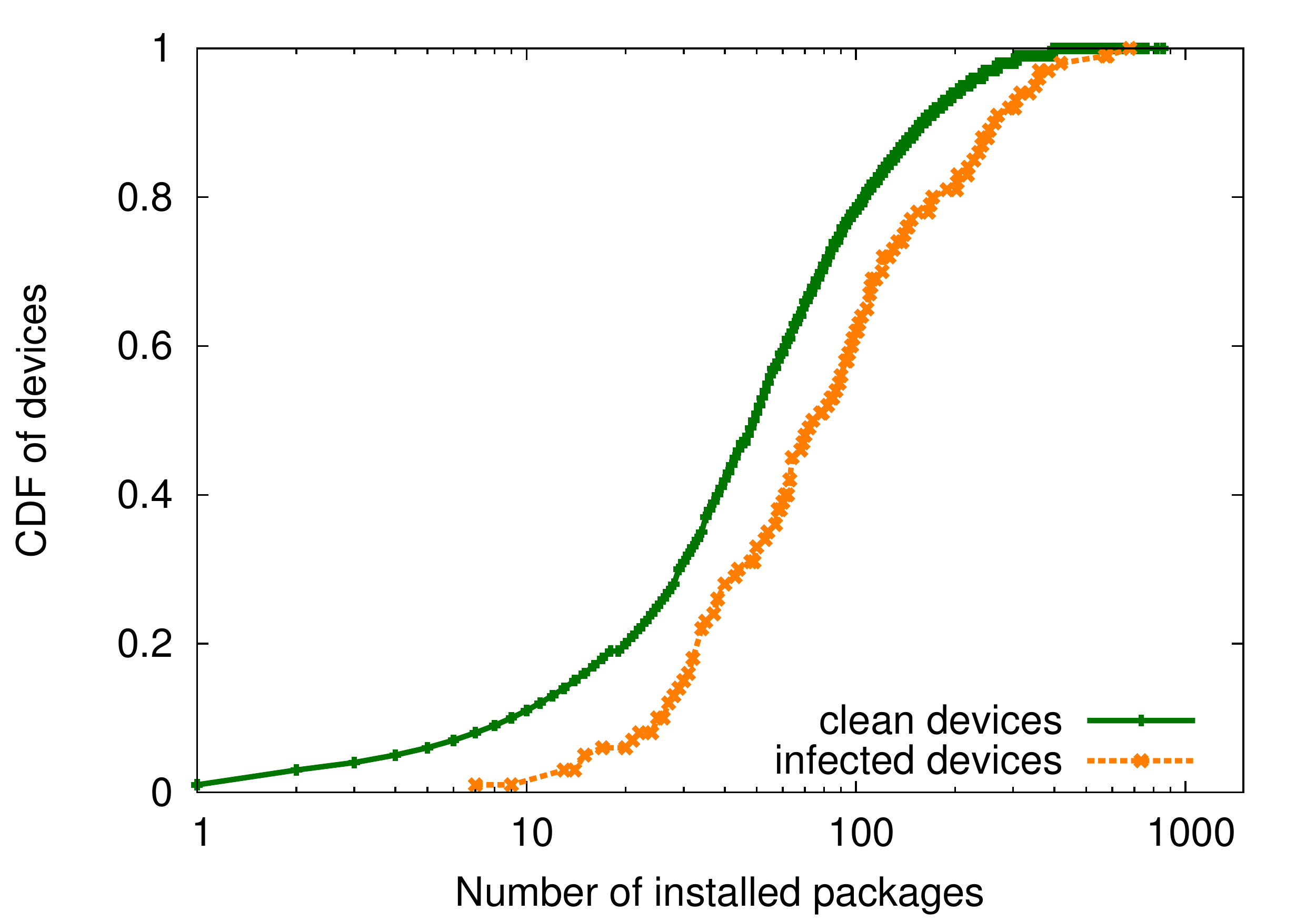}
    \subcaption{ McAfee dataset}
    \label{fig:package-device-dc-package-MC}
  \end{minipage}  
  \caption{Distribution of number of installed packages. \label{fig:pkg-installed}}
\end{figure*}
%%%%

\subsection{Number of Installed Packages}
\label{subsec:predictions:pkg-number}
%%%%
\begin{table}
\centering
\small{
\begin{tabular}{|l|c c|c c|}
\hline
& \multicolumn{2}{c|}{\textbf{Mobile Sandbox}} & \multicolumn{2}{c|}{\textbf{McAfee}} \\
\hline
\textit{Statistic} & \textit{Infected} & \textit{Clean} & \textit{Infected} & \textit{Clean} \\
\hline
\multicolumn{5}{c}{\textbf{All Models and OS Versions}} \\
\hline
Mean & 111 & 78.3  & 	 136 & 78.3  \\ 
Median & 85.5 & 53 & 	 97.5 & 53 \\ 
Wilcoxon & \multicolumn{2}{c|}{p < 0.01, Z=-4.57} & 	 \multicolumn{2}{c|}{p < 0.01, Z=-6.33} \\ 
\hline
\multicolumn{5}{c}{With 95th percentile high outlier removal} \\ 
\hline
Mean & 93.6 & 65.9 & 	 115 & 65.8 \\ 
Median & 83 & 50 & 	 93 & 50 \\ 
Wilcoxon & \multicolumn{2}{c|}{p < 0.01, Z=-4.83} & 	 \multicolumn{2}{c|}{p < 0.01, Z=-6.65} \\ 
\hline
\multicolumn{5}{c}{\textbf{Matching Models and OS Versions}} \\
\hline
Mean & 111 & 101  & 	 136 & 105  \\ 
Median & 85.5 & 80 & 	 97.5 & 79 \\ 
Wilcoxon & \multicolumn{2}{c|}{p=0.206, Z=-1.27} & 	 \multicolumn{2}{c|}{p < 0.01, Z=-2.97} \\ 
\hline
\multicolumn{5}{c}{With 95th percentile high outlier removal} \\
\hline
Mean & 93.6 & 88.3 & 	 115 & 90.7 \\ 
Median & 83 & 73 & 	 93 & 72 \\ 
Wilcoxon & \multicolumn{2}{c|}{p=0.198, Z=-1.29} & 	 \multicolumn{2}{c|}{p < 0.01, Z=-3.09} \\ 
\hline
\end{tabular}
}
\caption{Statistics on applications installed.\label{tbl:appresults}}
\end{table}
%%%%%%%%%%%%
We did similar outlier removal to applications as to battery life data.
We removed 484 outliers from clean and 6 from infected application data for Mobile Sandbox,
and 640 outliers from clean and 6 from infected application
data for McAfee.

The average number of applications observed on the infected devices
($93$, median $83$ for Mobile Sandbox / $115$, median $93$ for McAfee) was higher than the average number of applications on the clean devices
$(88$, median $73$ for Mobile Sandbox / $90$, median $72$ for McAfee) during our observation period. This would match with the
intuition that every newly installed application is an opportunity for
infection and that users who install and run more applications would therefore
be more likely to become infected. To assess this hypothesis more
rigorously, we used the Wilcoxon rank sum test to compare the number of installed
packages between infected and clean devices. The
difference was statistically significant for McAfee (Z = -3.086946, p < 0.01) and marginally significant for Mobile Sandbox (Z = -1.287166, p = 0.1980).
 
This is also illustrated in Figure~\ref{fig:pkg-installed}, which
shows the distributions of the number of installed packages for the
two device groups.  The detailed statistics are shown in
Table~\ref{tbl:appresults}.
\fi

\subsection{Applications Used on a Device}
\label{subsec:predictions:applications}

%% Our results indicate that the applications used on a device could be
%% used to detect likelihood of malware infection.  In the case of new
%% malware prediction, we can identify infected devices correctly $3.5$
%% times better than the baseline, random chance. For undetected malware
%% prediction, our approach is $3.8$ times better.

The density of malware in different application stores tends to vary
considerably, with some stores having a high malware incidence
rate~\cite{yajin_zhou_hey_2012}. The set of applications used on a
device can serve as a (weak) proxy for the application stores used by
the user of the device, thus potentially providing information about
the device's susceptibility to malware. Cross-application promotions
and on-device advertising are other factors that can affect
susceptibility to malware infections. As the next step of analysis, we
examined how much information about malware infection the applications
run on the device can provide.

We conduct our investigation by considering malware detection as a
classification task where the goal is to classify the device as
infected or clean using the set of applications used on the
device as the input feature. As discussed earlier, each anti-malware
vendor may have their own proprietary algorithm to classify
applications as malware or not.  Therefore, we report our detection
experiments separately for each malware dataset.

Analogously to, e.g., some spam filters, we rely on (Multinomial)
Na{\"i}ve Bayes classifiers in our experiments. Despite making a strong
independence assumption, Na{\"i}ve Bayes classifier is a powerful and
computationally efficient tool for analyzing sparse, large-scale
data. As the input features to the classifier we consider
bag-of-applications vectors, i.e., sparse binary vectors containing
value one for each application run on the device and the value zero
otherwise.  If a device contains an application that is present within
the list of known malware applications, we label the device as
infected. Otherwise, the device is labeled as clean.

As we are interested in looking at associations between malware and
other applications, and as anti-malware tools would easily detect
known malware applications, all applications known as malware were
removed from the data before carrying out the classification
experiments (they are used only to label devices). Since our data has
been collected over a period of multiple months, we also removed all
applications that corresponded to earlier (or newer) versions of one
of the malware applications (i.e., which have the same devcert and
package key, but different version code) to avoid any potential
temporal effects.

We start with a simple cross-validation experiment (detecting
infection by known malware) and then describe two variants of
detecting infection by new malware.  Finally we report
on a real-life scenario of detecting infection by previously
undetected malware.  In each experiment, the baseline for detection
is the success rate of finding an infected device by randomly
selecting a subset of supposedly clean devices.

\subsubsection{Cross-Validation}
\label{subsubsec:predictions:applications:cross}

We used stratified $5$-fold cross-validation. Accordingly, the set of devices was
partitioned into five subsets, each having a similar class distribution as
the entire dataset. Four of the subsets were used to train the
classifier and the remaining subset was used for testing. This process
was repeated five times so that each subset served once as the test
set.  Classification results were aggregated over the five folds. The
results are shown in 
\ifwww
Table~\ref{tbl:infdetect} (lines 1-2).
\else
Tables~\ref{tbl:5cv_MB} and
\ref{tbl:5cv_MC}.
\fi

The results clearly indicate the difficulty in accurately distinguishing malware
infections solely on the basis of applications run
on the device. Only a small number of the actually infected devices were
correctly identified, however, the classification algorithm also made
relatively few false detections (approximately 1000 devices out of
over $55,000$).  The precision is significantly better than
baseline (1.37\% and 1.07\%), random chance (it is also comparable to
what some anti-virus tools provide for these
applications~\cite{zhou_dissecting_2012}). More importantly, considering
the relatively low level of false positives, the results indicate that
applications run on the device could potentially be used as one of the
features for detecting likelihood of malware infection.

\ifwww
\else
%%%%%%%
\begin{table*}[htbp]
    \centering   
    \begin{minipage}{0.35\textwidth}      
        \begin{tabular}{|c | c | c | c |}
          \hhline{~~|--|}
          \multicolumn{2}{c |}{} & \multicolumn{2}{ c |}{\cellcolor{mygray} Detected class} \\
          \hhline{~~|-|-|}
          \multicolumn{2}{c |}{} & \multicolumn{1}{ c |}{ \cellcolor{mygray} Infected } & \cellcolor{mygray} Clean \\
          \hline
          \cellcolor{mygray} Actual & \cellcolor{mygray} Infected &  14  & 140 \\
          \hhline{|>{\arrayrulecolor{mygray}}->{\arrayrulecolor{black}}|-|-|-|}
          %\hhline{|{\arrayrulecolor{mygray}}-|-|-|-|}
          \cellcolor{mygray} class & \cellcolor{mygray} Clean   & 1,008 & 54,116 \\
          \hline
        \end{tabular}
        \subcaption{Mobile Sandbox, cross-validation, baseline 0.28\%,
          precision 1.37\%, 4.9X improvement. \label{tbl:5cv_MB}}     
    \end{minipage}
    %%%
    \begin{minipage}{0.45\textwidth}
      \begin{adjustwidth}{1.5cm}{}      
        \begin{tabular}{|c | c | c | c |}
          \hhline{~~|--|}
          \multicolumn{2}{c |}{} & \multicolumn{2}{ c |}{\cellcolor{mygray} Detected class} \\
          \hhline{~~|-|-|}
          \multicolumn{2}{c |}{} & \multicolumn{1}{ c |}{ \cellcolor{mygray} Infected } & \cellcolor{mygray} Clean \\
          \hline
          \cellcolor{mygray} Actual & \cellcolor{mygray} Infected &  11  & 133 \\
          \hhline{|>{\arrayrulecolor{mygray}}->{\arrayrulecolor{black}}|-|-|-|}
          %\hhline{|{\arrayrulecolor{mygray}}-|-|-|-|}
          \cellcolor{mygray} class & \cellcolor{mygray} Clean   & 1,020 & 54,114 \\
          \hline
        \end{tabular}    
        \subcaption{McAfee, cross-validation, baseline 0.26\%, precision 1.07\%, 4.1X improvement.\label{tbl:5cv_MC}}      
      \end{adjustwidth}
        \end{minipage}

%\vspace{1em}

     \begin{minipage}{0.35\textwidth}      
        \begin{tabular}{|c | c | c | c |}
          \hhline{~~|--|}
          \multicolumn{2}{c |}{} & \multicolumn{2}{ c |}{\cellcolor{mygray} Detected class} \\
          \hhline{~~|-|-|}
          \multicolumn{2}{c |}{} & \multicolumn{1}{ c |}{ \cellcolor{mygray} Infected } & \cellcolor{mygray} Clean \\
          \hline
          \cellcolor{mygray} Actual & \cellcolor{mygray} Infected &  53  & 687 \\
          \hhline{|>{\arrayrulecolor{mygray}}->{\arrayrulecolor{black}}|-|-|-|}
          %\hhline{|{\arrayrulecolor{mygray}}-|-|-|-|}
          \cellcolor{mygray} class & \cellcolor{mygray} Clean   & 5,100 & 267,450 \\
          \hline
        \end{tabular}
        \subcaption{Mobile Sandbox, new malware, baseline 0.27\%, precision 1.03\%, 3.8X improvement. \label{tbl:new_MB}} 
    \end{minipage}
    %% no blank line here
    \begin{minipage}{0.45\textwidth}
      \begin{adjustwidth}{1.5cm}{}      
        \begin{tabular}{|c | c | c | c |}
          \hhline{~~|--|}
          \multicolumn{2}{c |}{} & \multicolumn{2}{ c |}{\cellcolor{mygray} Detected class} \\
          \hhline{~~|-|-|}
          \multicolumn{2}{c |}{} & \multicolumn{1}{ c |}{ \cellcolor{mygray} Infected } & \cellcolor{mygray} Clean \\
          \hline
          \cellcolor{mygray} Actual & \cellcolor{mygray} Infected &  40  & 680 \\
          \hhline{|>{\arrayrulecolor{mygray}}->{\arrayrulecolor{black}}|-|-|-|}
          %\hhline{|{\arrayrulecolor{mygray}}-|-|-|-|}
          \cellcolor{mygray} class & \cellcolor{mygray} Clean   & 5,146 & 267,654 \\
          \hline
        \end{tabular}    
        \subcaption{McAfee, new malware, baseline 0.26\%, precision 0.77\%, 2.9X improvement. \label{tbl:new_MC}}      
      \end{adjustwidth}
   \end{minipage}

%\vspace{1em}

   \begin{minipage}{0.35\textwidth}      
        \begin{tabular}{|c | c | c | c |}
          \hhline{~~|--|}
          \multicolumn{2}{c |}{} & \multicolumn{2}{ c |}{\cellcolor{mygray} Detected class} \\
          \hhline{~~|-|-|}
          \multicolumn{2}{c |}{} & \multicolumn{1}{ c |}{ \cellcolor{mygray} Infected } & \cellcolor{mygray} Clean \\
          \hline
          \cellcolor{mygray} Actual & \cellcolor{mygray} Infected &  5  & 132 \\
          \hhline{|>{\arrayrulecolor{mygray}}->{\arrayrulecolor{black}}|-|-|-|}
          %\hhline{|{\arrayrulecolor{mygray}}-|-|-|-|}
          \cellcolor{mygray} class & \cellcolor{mygray} Clean   & 5,098 & 270,572 \\
          \hline
        \end{tabular}
        \subcaption{Mobile Sandbox, undetected malware, baseline 0.05\%, precision 0.10\%, 2.0X improvement. \label{tbl:undetected_MB}}
    \end{minipage}
    %%%
    \begin{minipage}{0.45\textwidth}
      \begin{adjustwidth}{1.5cm}{}      
        \begin{tabular}{|c | c | c | c |}
          \hhline{~~|--|}
          \multicolumn{2}{c |}{} & \multicolumn{2}{ c |}{\cellcolor{mygray} Detected class} \\
          \hhline{~~|-|-|}
          \multicolumn{2}{c |}{} & \multicolumn{1}{ c |}{ \cellcolor{mygray} Infected } & \cellcolor{mygray} Clean \\
          \hline
          \cellcolor{mygray} Actual & \cellcolor{mygray} Infected &  7  & 120 \\
          \hhline{|>{\arrayrulecolor{mygray}}->{\arrayrulecolor{black}}|-|-|-|}
          %\hhline{|{\arrayrulecolor{mygray}}-|-|-|-|}
          \cellcolor{mygray} class & \cellcolor{mygray} Clean   & 5,044 & 270,721 \\
          \hline
        \end{tabular}    
        \subcaption{McAfee, undetected malware, baseline 0.05\%, precision 0.14\%, 3.0X improvement. \label{tbl:undetected_MC}}      
      \end{adjustwidth}
    \end{minipage}

%\vspace{1em}

   \begin{minipage}{0.35\textwidth}      
        \begin{tabular}{|c | c | c | c |}
          \hhline{~~|--|}
          \multicolumn{2}{c |}{} & \multicolumn{2}{ c |}{\cellcolor{mygray} Detected class} \\
          \hhline{~~|-|-|}
          \multicolumn{2}{c |}{} & \multicolumn{1}{ c |}{ \cellcolor{mygray} Infected } & \cellcolor{mygray} Clean \\
          \hline
          \cellcolor{mygray} Actual & \cellcolor{mygray} Infected &  4  & 71 \\
          \hhline{|>{\arrayrulecolor{mygray}}->{\arrayrulecolor{black}}|-|-|-|}
          %\hhline{|{\arrayrulecolor{mygray}}-|-|-|-|}
          \cellcolor{mygray} class & \cellcolor{mygray} Clean   & 609 & 54,515 \\
          \hline
        \end{tabular}
        \subcaption{Mobile Sandbox, train with ``original'' set,
          detect infection wrt ``new'' set, baseline 0.14\%,
          precision 0.65\%, 4.8X improvement. \label{tbl:real_MB}}
    \end{minipage}
    %%%
    \begin{minipage}{0.45\textwidth}
      \begin{adjustwidth}{1.5cm}{}      
        \begin{tabular}{|c | c | c | c |}
          \hhline{~~|--|}
          \multicolumn{2}{c |}{} & \multicolumn{2}{ c |}{\cellcolor{mygray} Detected class} \\
          \hhline{~~|-|-|}
          \multicolumn{2}{c |}{} & \multicolumn{1}{ c |}{ \cellcolor{mygray} Infected } & \cellcolor{mygray} Clean \\
          \hline
          \cellcolor{mygray} Actual & \cellcolor{mygray} Infected &  4  & 112 \\
          \hhline{|>{\arrayrulecolor{mygray}}->{\arrayrulecolor{black}}|-|-|-|}
          %\hhline{|{\arrayrulecolor{mygray}}-|-|-|-|}
          \cellcolor{mygray} class & \cellcolor{mygray} Clean   & 407 & 54,727 \\
          \hline
        \end{tabular}    
        \subcaption{McAfee, train with ``original'' set,
          detect infection w.r.t ``new'' set, baseline 0.21\%, precision 0.97\%, 4.6X improvement. \label{tbl:real_MC}}      
      \end{adjustwidth}
    \end{minipage}
%%%
    \caption{Detection of infection based on the set of applications
      on a device.}    
\end{table*}
%%%%%%%%%%%%%
\fi

\subsubsection{Infection by New Malware}
\label{subsubsec:predictions:applications:new}

Next we evaluate the potential of using the set of applications used
on a device as an {\predictor}  for infection by \textit{new, previously
  unknown} malware.  To do this, we partitioned the set of clean devices
into a training set consisting of $80\%$ and a test set consisting of
20\%, chosen randomly. We partitioned the malware five ways, according to the number of infected devices.
These
groups therefore correspond to roughly equal number of infected
devices. 

In each run (of a total of five runs), four groups were used as ``known
malware'' and the devices they infected were combined with the
training set ($80\%$ clean devices).  The remaining group was used as
``unknown malware'' and its infected devices combined with the test
set ($20\%$ clean devices).
%% In every run of a total five runs, four groups were used as
%% ``known malware'' and the remaining group was used as ``unknown
%% malware''.  The devices infected by ``known malware'' were combined
%% with the training set ($80\%$ clean devices) and the resulting set was
%% used to train the prediction model. The devices infected by ``unknown
%% malware'' were combined with the test set ($20\%$ clean devices) and
%% the resulting set was used as the test set for the model.

No two devices infected by the same malware application appear in both
(training and test) sets in the same run to ensure that malware
applications for testing are truly unknown. In line with our other
experiment, we also removed all application features corresponding to
newer or older versions of a malware application. Lastly, to mitigate
the influence of random partitioning, we repeated the entire experiment
five times and report the summed results (the aggregation was done
over 25 runs). Results are reported in 
\ifwww
Table~\ref{tbl:infdetect} (lines 3-4).
\else
Tables~\ref{tbl:new_MB} and
\ref{tbl:new_MC}.
\fi

The classification results for detecting infection
by new malware are 3.8 (Mobile Sandbox) and 2.9 (McAfee) times better
than baseline.

In the experiment, all the devices that had an application from the
set of ``undetected malware'' were always assigned to the test set.
While this is reasonable for truly new, unknown malware, in reality,
it is possible that undetected malware was present on some devices in
the set of devices used to train the model (naturally those devices
would have been classified as ``clean'' at the time of training). We
next investigate the potential for detecting infection by previously
undetected malware.

\subsubsection{Infection by Previously Undetected Malware}
\label{subsubsec:predictions:applications:undetected}

Infection by previously undetected malware (including new as well as
old but previously undetected malware) may also be detectable using
the same classification approach.  To evaluate this possibility, we
ran a new experiment.
%In this experiment, we evaluate the possibility of detecting device 
%infection by previously undetected malware (including new as well as old but previously
%undetected malware). 
We partitioned the
set of \textit{all} devices randomly into two sets: a training set
containing 80\% of the devices and a test set containing the remaining
20\%.  Next, we partitioned the set of malware applications five ways
according to the number of infected devices. 
In each run, only the
malware applications from four malware sets (i.e., ``known malware''
sets) were used to label ``infected'' devices in the training set.
Any device in the training set that contains an application from the
remaining malware set (i.e., ``undetected malware'') was labeled as
``clean'' to reflect the fact that at the time of training such
devices were not known to be infected.  We moved any device in the
test set that is infected by ``known malware'' to the training set.
To minimize the effect of random partitioning, as before, we repeated
the entire experiment five times, with five runs in each round, and report
the summed results of 25 runs in 
\ifwww
Table~\ref{tbl:infdetect} (lines 5-6).
\else
Tables~\ref{tbl:undetected_MB}
and \ref{tbl:undetected_MC}.
\fi 
The results are 2 (Mobile Sandbox) and 3
(McAfee) times better than the baseline.

\subsubsection{Detection of `Real life'' Infections}
\label{subsubsec:predictions:applications:reallife}

So far, we simulated ``new'' or ``previously undetected'' malware by
dividing our malware datasets randomly.  For each of our malware
datasets we received a first version (the ``original'' set) in March
2013 and an updated version (the ``updated'' set) in subsequently (in
September 2013 for Mobile Sandbox and November 2013 for McAfee).  This
allowed us to validate our model for detection of infection by
previously undetected malware under ``real life'' conditions.  Let us
denote the difference between the updated set and the original set as
the ``new'' set.  We labeled the full set of devices using the
original malware set, trained our model using the resulting set and
used the set of all ``clean'' devices (with respect to the original
set) as the test set to detect infection. We compared the results
with respect to infections labeled by the ``new'' set.
% In previous experiments, we generated randomly data for training and
% testing based on chosen division, i.e. 80\% and 20\%. To validate the
% approach, we also apply the same method on our ``real-life''
% dataset. Our malware dataset is updated at different times. The
% initial malware datasets were collected in March, 2013. In September,
% 2013 (for Mobile Sandbox) and November, 2013 (for McAfee), we continue to
% have new updates for those malware sets. To simulate the case where an
% AV vendor have in hand the known malware set collected in March, and
% the new malware set collected after that were not yet known, we try
% the prediction algorithm to detect infected devices with these new
% malware. That means at the training period, only devices that have
% applications appearing in known malware sets are labeled as
% infected. The devices that have applications appearing in unknown
% malware sets are still labeled as clean. All devices in \caratapp{}
% were used for training and testing. However, in the dataset for
% testing, all infected devices by known malware are removed, and only
%devices infected by unknown malware are kept for prediction. 
The results are shown in 
\ifwww
Table~\ref{tbl:infdetect} (lines 7-8).
\else
Tables~\ref{tbl:real_MB} and
\ref{tbl:real_MC}.
\fi 
The detection performance is 4.8 (Mobile Sandbox)
and 4.6 (McAfee) times better than the baseline of random selection.

%\newpage
\section{Discussion}
\label{sec:discussion}
\noindent\textbf{Infection rates}:
The infection rates we find are nowhere as high as the more alarmist
claims from some in the anti-malware
industry~\cite{nqmobile_mobile_2013_wurl}. But even our most
conservative estimates (\McAfeeinfectionrate and \MobSandinfectionrate) are still (statistically)
significantly\footnote{A two-sample proportions test
(Chi-squared) verified that the difference is statistically highly
significant (Mobile Sandbox: $\chi^2$=44132.24, df=1, p<0.001, and McAfee:
$\chi^2$=38657.59, df=1, p<0.001.)} higher than what was reported by
Lever~et al.~\cite{charles_lever_core_2013}.  There are a
number of possible explanations:

\begin{itemize}

\item Lever~et al.~looked at devices in the United States only.  It is
  believed that the prevalence of mobile malware infection is much
  higher in China and Russia. We cannot be certain about how many of
  the infected devices we found are located in those regions.  However,
  as we discussed in Section~\ref{subsec:predictions:language}, we can estimate a lower bound of
  at least
  13 infected devices (0.02\%) as being in the United States.
  
\item Lever~et al.~identified infected devices by examining device DNS requests
  for tainted hosts. Not all malware generates DNS requests (e.g.,
  malware that aims to send premium SMS messages may not bother with
  DNS requests). Malware authors presumably change their data
  collection and command and control addresses frequently or may have
  used hardwired IP addresses. Both of these factors may have led to
  an underestimation of the infection rate.

\end{itemize}

Our conservative malware infection rate estimates are closer to
Google's recent estimate,
0.12\%~\cite{google_malware_estimate_2013}. However, Google's estimate
refers to the percentage of application installations that they have
marked as potentially harmful, while ours is the percentage of
infected devices in a community of mostly clean devices.
%% Asokan: commenting for the reason indicated in e-mail to Eemil.
%% If Google is counting application installations, then
%% one-malware-per-device will significantly.
%% Therefore
%% Google's number estimates the amount of infected devices only if
%% a single device only installs one potentially harmful application.
Also, Google affects the result by warning the user at installation time,
while \caratapp{} collects the information on running applications
after installation. Further, installation of an application does not guarantee
that it will ever be run.

\noindent{\textbf{Detecting infection}}: The results of the
classification experiments in
Section~\ref{subsec:predictions:applications} demonstrate that the
set of applications used on the device is a potential source of
information for detecting malware applications.  The detection
techniques discussed in 
\ifwww
Section~\ref{subsec:predictions:applications},
\else 
Sections~\ref{subsubsec:predictions:applications:new},
\ref{subsubsec:predictions:applications:undetected}, and
\ref{subsubsec:predictions:applications:reallife}
\fi
are not intended to replace the standard anti-malware scanning for
infection by \textit{known} malware.  Rather they can complement
standard anti-malware tools in a number of ways.  We foresee two ways
in which early warning could be used:
\begin{itemize}
\item \emph{Search for previously undetected malware}: An anti-malware
  vendor can, after doing the standard scanning for malware using
  known malware signatures, apply our detection technique to the list
  of applications used on a device to determine if the device falls in
  the ``vulnerable'' class and inform the vendor of the tool if it
  does.  The vendor can then apply expensive analysis techniques on
  all the applications in vulnerable devices.  Without such a
  detection mechanism, the vendor would have had to take a random
  sample of devices.  
\item \emph{Training enterprise users}: Consider an enterprise that
  has a small budget for training users on good application hygiene.  Rather
  than selecting trainees at random, the enterprise administrators
  could target the training towards users of the ``vulnerable'' class
  of devices.
\end{itemize}
For example, in the case shown in 
\ifwww
Table~\ref{tbl:infdetect} (line 8), 
\else
Table~\ref{tbl:real_MC}, 
\fi
the success
rate for finding infection by previously unknown malware by taking a
random sample of devices is 0.21\%.  Our detection mechanism results
in an almost five-fold improvement (precision=0.97\%).  It is
important to note that this improvement comes at virtually no cost
because the instrumentation needed to collect the data is extremely
lightweight. Our detections were based only on the set of
applications used on a device, where each application is identified
only by the \dcpv tuple. Measuring this set of active applications
intermittently using \caratapp{} incurs negligible performance or
battery overhead (literally below the precision of our hardware
instrumentation).

The effectiveness of the detections could be improved with more data,
and with additional features, such as battery consumption and the amount
and extent of permissions required by the application.  More effective
detection techniques can lead to other early warning applications.
For example, \caratapp{} or a standalone mobile security application
on the device might visualize the detection as a traffic light
indicator of ``threat level''.

\noindent\textbf{Predicting expected time before infection}:
\label{subsec:predictions:time}
Our approach is well--suited to answering the question of the
\emph{expected time before infection}. This is an interesting and
important question and a solution would provide a much needed temporal
risk measure for end users. In this section, we outline our simple
solution for this problem that will be verified in our future work. In
order to fully develop and deploy a solution, we need the application
specific time-to-infection distributions from devices. 
%% At the moment, we do not have sufficient empirical data to determine
%% these distributions.  Since our data collection period lasted only for
%% about 40 days, we found 21 devices whose status changed from ``clean''
%% to ``infected'' \textit{during} the data collection process. 
We expect to see more such cases as we continue the data collection
over the next few months.  Once we have enough such cases, how can we
predict the expected time to infection?

Our proposed solution is based on the application--specific
time-to-infection distributions. For a given application and device,
we record the time the application has been installed on a clean
device until the device became infected. This time is zero for
applications installed on already infected devices and infinite for
clean devices. Thus we obtain per application distributions of the
time-to-infection.

Based on the application specific time-to-infection distributions we
then determine the device specific time-to-infection. This is
performed by considering each application on the given clean device
and then determining the expected minimum time-to-infection based on
the application specific distributions.  We believe that the minimum
time is a reasonable starting point for assessing the risk to the
user.

The three important parts of the process are the summary statistic for
the application specific distributions, application correlations, and
the determination of the expected minimum time.

Our initial solution uses the median to summarize the application
specific distributions. The median is a robust statistic and it copes
well with outliers. We take correlations between applications into
account by considering also subsets of applications. The motivation is
that the summary statistic may hide interesting patterns pertaining to
certain groups of applications that together significantly reduce the
time-to-infection. A group is included in the analysis if its time to
infection value is lower than the respective values of its elements.
Given the application and group statistics, it is then easy to choose
the minimum value.

\noindent\textbf{Energy consumption}:
We found a marginally significant difference between infected and
clean devices in terms of remaining battery life. Our results show
that malware reduces the remaining operating time of the devices by an
average of $1.3$ hours (Mobile Sandbox) and $0.4$ hours (McAfee). We
are incentivized to detect and remove malware in order to conserve
energy. In addition, we can use this observation to detect
malware. The combination of installed and running applications with
the energy consumption data make for a new way to assess the risk of
having malware.  Crowdsourcing of malware detection has been proposed
before~\cite{burguera_crowdroid:_2011}, however, our approach is
unique because it only uses knowledge of the installed, running
applications and the energy consumption, making it non-intrusive and
lightweight. The combination of these two features---applications and
energy---appears to be a promising avenue for future research.

%ADAM I left this here in case you want to integrate it
%(this seems to be relevant prior work?
%\cite{Kim:2008:DEA:1378600.1378627}).

\noindent\textbf{Reliable malware detection}:
A typical anti-virus tool performs extensive analysis of a package on
a device in order to determine if it is malware.  We did not have this
luxury because we wanted to minimize the overhead we add to
\caratapp{}. Consequently, we resorted to identifying malware by
comparing reliable identifiers for packages (e.g., \dcpv
tuples) with those in known malware datasets. This may underestimate
the incidence of malware in the sense that it will not detect any
previously unknown malware, just as any other signature-based malware
detection mechanism.
% ADAM doesn't this sentence above call into question all of our
% infection rate numbers?
% ASOKAN: It under-estimates in the same way as other signature-based
% AV tools miss zero-day malware
Despite this limitation, our results 
provide new, and more accurate, information about malware
  infection rates which suggests that mobile malware infection may be
  more widespread than previous rigorous independent estimates.

  Our detection techniques are independent of the method used to
  decide if a device is infected or not. Consequently, their efficacy
  will be improved when they are used together with better techniques
  for identifying malware.

\noindent\textbf{Limitations}:
In our \caratapp{} dataset, we did not make use of the time
information associated with \caratapp{} samples. It is possible that users
infected by malware go on to install anti-malware or other performance
analysis tool in order to troubleshoot.
Since our analysis technique is based
on applications occurring together on infected and clean devices,
it may incorrectly infer that the presence of such applications is an
indicator of potential infection. Removing such applications from the
analysis would be one way to address this problem.
For privacy reasons, \caratapp does not collect any demographic data
about its users. Consequently, we cannot be certain that the
\caratapp dataset corresponds to a representative sample of Android
users in general besides the geographical distribution information we
presented in Section~\ref{subsec:background:carat}.

\section{Related Work}
\label{sec:relatedwork}
Work by Lever~et al.~\cite{charles_lever_core_2013} was the first (and
until now, to the best of our knowledge, the only) public, independent
study of mobile malware infection rates. They used large datasets
consisting of DNS requests made by the customers of a US-based
Internet service provider and a cellular carrier and tried to identify
infected mobile devices based on DNS requests to known tainted
hostnames. Our work, in contrast, uses data collected directly from
the mobile devices.

Analyzing mobile malware has been an active research area. Felt~et al.~\cite{felt_survey_2011} presented one of the first surveys of
malware on three different mobile platforms. Zhou and Jiang
\cite{zhou_dissecting_2012} provide a detailed, systematic analysis and
classification of Android malware based on a large set of malware
samples. Some researchers have explored collaborative techniques for
malware detection~\cite{burguera_crowdroid:_2011,yang_enhancing_2011}.
Our work differs from these in that rather than detecting malware as
such, we use data collected from a large number of devices to 
%predict the impending arrival of malware on a device.
quantify the susceptibility of infection for a given device.
% ADAM why the "(some)"?; -- DONE (removed "(some)")

This paper proposes using proxy signals, like energy use or the set of
applications run on a device, to detect or predict infection. That
kind of approach bears a resemblance to anomaly detection, especially
in the intrusion detection field, which has a rich
history~\cite{Paxson:99,Sebring:88}.  Kim~et
al.~\cite{Kim:2008:DEA:1378600.1378627} proposed a power-aware malware
detection framework for mobile devices targeted for detecting battery
exhaustion malware. One paper suggests that energy is an insufficient
metric for detecting malware~\cite{hoffmannmobile}, but energy issues
have been successfully attributed to buggy and malicious
applications~\cite{Pathak:12a,caratSensys2013}.

% Removed neural network paper: Debar:92,
% Removed haystack, helps users detect intrusion manually: Smaha:88,
% Removed Vaccaro's rule-based system: ,Vaccaro:89

\iffalse
That kind of approach has been used before, also in the intrusion
detection field, for false-positive
suppression~\cite{Cuppens:02,Valdes:01}, cumulative
triggers~\cite{Huang:07}, alarm aggregation and
correlation~\cite{Bouloutas:94,Jakobson:93,Weaver:04}, and for
characterizing attack scenarios and causal flow \cite{Ning:02}. The
community has also been used to reduce per-client
overhead~\cite{Locasto:05} and increase statistical
confidence~\cite{Oliner:10}.  There are also uses of the community for
tasks other than detection, such as diagnosing problems by discovering
root causes \cite{Wang:04a} and preventing known exploits (e.g.,
sharing antibodies) \cite{Brumley:07,Costa:05,Newsome:06a}.
\fi
% Removed graph-based network attack detection: Staniford-chen:96,
%Remove Ullrich, url to internet storm center DShield.org

Motivated by the prohibitive cost of doing comprehensive malware
detection on mobile phones, Portokalidis~et al.~proposed Paranoid
Android~\cite{portokalidis_paranoid_2010} which maintains exact
virtual replicas of mobile devices on a server where the expensive
malware analysis is done.  Maintaining exact replicas may be
considered privacy-invasive~\cite{chandramohan_detection_2012}.  Also,
analyzing application permissions has been
done~\cite{chiapermissions2012}, but also safe applications often
request extensive permissions, making application permissions an
insufficient indicator of malware.

Our approach of using lightweight instrumentation to identify
potentially vulnerable devices can help reduce the impact of the
privacy concern. Our work uses data collected from a large number of
clients (sometimes called a {\em community}) to build statistical
models.  Much of the academic research in applying statistical
analysis and machine learning to the problem of mobile malware takes a
\emph{software-centric} view, focusing on analyzing software packages
to determine if they are malware. In contrast, we take a
\emph{device-centric} view, attempting to estimate the propensity of a
device for infection.  The closest prior work in this aspect was by
Wagner~et al.~\cite{Wagner2012socialbots} and subsequently Sumber and
Wald \cite{SW13_Blackhat} who took a similar approach in the context
of Twitter.  They use publicly visible characteristics of Twitter
users to predict their susceptibility to fall victim to social bots.

Finally, some recent activity focuses on collecting and collating
mobile applications (both malware and otherwise) and making them
available publicly in a systematic manner with well-designed
interfaces~\cite{barrera_understanding_2012,michael_spreitzenbarth_mobilesandbox:_2013}. These
have been extremely useful in our work.

\section{Conclusion}
\label{sec:conclusion}
In this paper, we addressed a gap in the research literature regarding
malware infection rates on mobile devices by direct measurement on
tens of thousands of mobile devices.  Our estimates for infection rate
of mobile malware, although small (\MobSandinfectionrate for Mobile
Sandbox and \McAfeeinfectionrate for McAfee), is still higher than
previous estimates.
%We also found evidence that the presence of mobile malware may be directly correlated with the number of applications installed and run by a user and inversely correlated with expected battery life.

We also investigated whether we can build models that can detect
malware infection based on the set of applications
currently installed on a device.  Although the precision and recall of
this initial detection attempt are not high, the approach can still
constitute one line of defense in a suite of techniques, especially
given that the data collection needed for the detection is extremely
lightweight. In particular, our models can be used by enterprise IT
administrators and anti-malware vendors to identify a small pool of
vulnerable devices, e.g., to deploy more expensive analysis techniques
on them or to provide training to their users.  In our experiments,
the precision of the model is up to five times better than the
baseline of random selection of devices.

There are several interesting directions to continue the work,
including the following:
\begin{itemize}
\item \emph{Using a larger dataset}: Repeating and improving our
  analysis using the larger \caratapp dataset.
\item \emph{Improving detection accuracy}: Using additional features
  and/or experimenting with better detection techniques to improve
  the precision and recall.
\item \emph{Predicting expected time to infection}: Validating our
  proposed approach for predicting expected time to infection using a
  larger dataset from \caratapp{}.
% We do this already.  Include graph in research report.
% \item \emph{Analyzing usage of anti-virus software}: Quantifying the
%   extent to which anti-virus software is used by the \caratapp{} user
%   base and studying their effectiveness.
  \item \emph{Energy vs.\ infection}: Investigating whether
    malware infection has an impact on the expected battery life and whether another proxy for
    energy use could be predictive or indicative of infection.
\end{itemize}
\ifwww
An extended version of this paper is available~\cite{TruongMobileMalware2013}.
\else
This research report is an extended version of a WWW 2014 paper~\cite{TruongMobileMalwareWWW2013}.
\fi
%We intend to make the \caratapp{} dataset used in this paper publicly available.

%

\section*{Acknowledgements} 

{\small We thank Igor Muttik, Michael Spreitzenbarth, Xuxian
  Jiang, and Yajin Zhou for giving us their respective malware
  datasets.  In particular, Michael and Igor responded promptly to
  several e-mail requests for clarifications additional information.
  We thank the anonymous reviewers for their valuable feedback.
  The work of Truong and Bhattacharya was supported by the
  Intel Institute for Collaborative Research in Secure Computing
  (ICRI-SC)}.

% For plain bibtex

% \ifwww
% %% new page only in WWW
% \newpage
% .
% \pagebreak
% \fi
%\balance
\bibliographystyle{abbrv}
\bibliography{full-paper}

% For biblatex
%\printbibliography

 \end{document}